\documentclass[12pt]{article}
\usepackage{graphicx,amssymb,amsmath,amsthm,cite}

\usepackage{color,multirow}
\usepackage{verbatim}
\newcommand{\la}{\langle}
\newcommand{\ra}{\rangle}

\newcommand{\Spann}{{\mbox{\rm{span}}}}
\newcommand{\til}{\tilde}

\newcommand{\vc}{\mathbf{c}}

\newtheorem{remark}{Remark}

\def\be{\begin{equation}}
\def\ee{\end{equation}}
\def\ben{\begin{eqnarray}}
\def\een{\end{eqnarray}}

\def\op{\hat{\mathrm{P}}}
\def\orth{\widehat{\mathrm{Orth}}}
\def\ll{{\ell}}

\DeclareMathOperator*{\nul}{null}

\def\PSNR{\mathrm{PSNR}}
\def\MSE{\mathrm{MSE}}
\def\bit{\mathrm{b}}
\def\key{\mathbf{key}}
\def\seed{\mathbf{seed}}
\def\Nx{N_x}
\def\Ny{N_y}
\def\Nxy{\Nx \times \Ny}
\def\Mx{M_x}
\def\My{M_y}
\def\dx{{D}^x}
\def\dy{{D}^y}

\def\It{{\til{I}}}
\def\Nt{{\til{N}}}
\def\Lt{{\til{L}}}

\def\Qt{{\til{H}}}
\def\Yt{{\til{Y}}}
\def\Xt{{\til{X}}}
\def\Ft{{\til{F}}}
\def\Gt{\til{G}}
\def\Ut{{\til{U}}}

\def\htt{{\til{h}}}
\def\Dx{\mathcal{D}^x}
\def\Dy{\mathcal{D}^y}
\def\D{\mathcal{D}}
\def\R{\mathbb{R}}
\def\N{\mathbb{N}}
\def\V{\mathbb{V}}
\def\F{\mathrm{F}}

\title{Self Contained Encrypted Image Folding}
\author{L. Rebollo-Neira and J. Bowley\\
Mathematics Department, Aston University\\
Birmingham, B4 7ET, UK\\
\vspace{0.1cm}\\
A. G. Constantinides\\
Department of Electrical and Electronic Engineering\\
Imperial College, UK\\
Exhibition Road, London SW7 2BT, UK\\
\vspace{0.1cm}\\
A. Plastino\\
IFLP-CCT-Conicet\\
National University of La Plata\\
CC 727, 1900 La Plata, Argentina}
\begin{document}
\maketitle
\begin{abstract}
The recently introduced approach for Encrypted Image 
Folding is generalized to make it self-contained. 
The goal is achieved by enlarging the folded 
image so as to embed all the necessary information
for the image recovery. The need for  
extra size is somewhat compensated by considering 
a transformation with higher folding capacity.
Numerical examples show that the size of 
the resulting {\em{cipher image}} may be significantly 
smaller than the {\em{plain text}} one.
The implementation of the approach is further
extended to deal also with color images.
\end{abstract}
\section{Introduction}
As cameras and digital scanners of very high
resolution are becoming widely available,
use of high resolution digital images is becoming part 
of everyday life.
From a mathematical standpoint a digital image is 
 a $2$D data array, say  $I \in \R^{\Nxy}$. Each
data point is referred to as a
pixel. For a gray level image, each pixel  
 is represented with an intensity value $I$.
For an RGB representation of a color image,
each pixel consists of a color triple
$(I_R, I_G, I_B)$ representing the intensity of the red,
green and blue components, respectively.

The array of pixels used to represent a high resolution
digital image is expected to be huge.  
Obviously storage and  transmission of this raw
data is impractical.
Consequently, a reduction in data  
dimensionality is essential. The process that creates a compact
data representation is called {\em{compression}}. Because of the
nature of its informational content compressing an image
usually involves special techniques. 
As opposed to binary files where a single bit error 
may destroy the whole piece of data, some distortion
is usually tolerable even when compressing high quality images.
This is because the visual perception of the image is more important than the exact pixel values.

The most frequently applied image compression techniques 
involve transform coding which has three main steps: 
i)Application of an invertible transform to the intensity image.
ii)Quantization of the transformed data.
iii)Bit-stream coding.

The familiar compression standard JPEG, for instance, 
implements step i) using the Discrete Cosine
Transform (DCT), while the more recent, 
JPEG2000, uses Discrete Wavelet Transform (DWT).

Another problem associated with the transmission of
digital images is security. It comprises
several aspects, including confidentiality and access control which are
addressed by encryption. This implies that only parties
holding decryption keys can access content of an image.
Conventional image encryption  
is based on techniques developed
for general data \cite{book1,book2}.
In principle generic encryption can be applied to a digital image
before or after compression. However, encryption before
compression would change the statistical properties of 
the image preventing compression from being applied 
successfully. 

On the other hand, as well as effecting the compression performance,
direct encryption of the compressed data results in a bit stream that is incompatible with the original image file format.
Less stringent schemes involve partial 
(or selective) encryption \cite{book1,book2}. 
However the security of these encryption systems 
is lower when compared to full encryption.

Enhancing security of conventional
compression/encryption techniques, using a
chaotic map at the bit-stream coding step,
is proposed in \cite{YW11,LWC12}.
However, for the most part, the line of research
for image encryption based on
Chaotic Cryptography \cite{Bap98,CMC04,MCL04,LSW05,
AKS07, XWL07, PZ08, ARA08, ALA10, SCY10, KL11}
has been developed to operate directly on the pixel$/$intensity 
representation of an image.
An interesting critical analysis of the research in this area
can be found in \cite{AAA11}. The  
connection between chaotic and conventional cryptography is
considered in \cite{MAD08}.

Chaos based image encryption takes advantage of 
the extreme sensitivity to initial
conditions of some dynamical systems, to control the
`confusion' of pixels in an intensity image.  

Thus, a chaotic method breaks the structure of 
the plain text image, producing a cipher image which
is no longer compressible by conventional transform coding techniques.
Hence, within the traditional chaos based framework for image encryption 
the problem of storage and  transmission of large images 
is currently unsolved.

An alternative framework, involving 
only mathematical operations on an intensity image, but 
addressing simultaneously the
problems of data reduction and encryption, 
has been recently introduced in \cite{BRN11}. The scheme is termed 
Encrypted Image Folding (EIF).
The first step of this new scheme differs from  
step i) in the above mentioned conventional compression scheme
in that, instead of using orthogonal transformations (e.g. DCT or 
DWT) the transformation is realized by means of
{\em{highly nonlinear approximation}} techniques. 
This increases the difficulty of the approximation 
process but at the same time renders significant improvement 
in the sparsity of the image representation.

Quantization and data reduction are achieved simultaneously
by embedding some of the transformed data into a section of the 
image.  Privacy is protected by 
granting access to the embedded data only 
to key holders.

The underlying principle of the proposed 
framework is very simple: Suppose that an  
image is given as an intensity array
$I \in \R^{\Nxy}$ 
and suppose also that, 
through a transformation $\hat{B}: \R^{\Nxy} \to 
\R^{K}$, one can approximate equivalent
information from an array $c \in \R^{K}$ 
obtained as $c= \hat{B} I$. If $K < \Nx \Ny$  by a 
considerable amount, $c$ is said to be a {\em{sparse 
representation}} of the image $I$. It follows then that 
a suitable transformation to achieve sparsity 
 should be rank deficient, with an associated null 
 space, $\nul(\hat{B})$, of large dimensionality. 
Such a transformation {\em{creates room
for storing covert information}}. Indeed, 
if one considers an element 
$F \in \nul(\hat{B})$ and adds it to the image, so as to 
create a new array
$G=I + F$, one obtains the identical representation
$\hat{B} G= \hat{B} I=c$.
The sparser the representation of an image, the larger the
null space of the associated transformation.
Consequently, the first part of this effort focuses
on the design of an effective transformation for this purpose. 
The transformation is  adaptively constructed by 
the greedy selection strategy 
called Orthogonal Matching Pursuit (OMP).

The viability of EIF, as proposed in \cite{BRN11}, 
stems from the possibility of processing a large image by 
dividing it into {\em{small blocks}}. This allows the representation 
of some of the blocks to be embedded into other blocks, realizing 
in that manner the folding of the image. However, the technique 
in \cite{BRN11} is not self-contained, because, in addition 
to the folded image, extra
information is required at the unfolding step, and 
that information depends on the image. 
In this Communication we propose to extend EIF so as to make it 
self-contained. We term such an extension Self-Contained 
Encrypted Image Folding (SCEIF), because all that  
is needed to successfully unfold the image 
 is the private key. This goal is 
accomplished by enlarging the folded image to create further 
space for the required information. The  need for extra size
is compensated by considering a transformation
with the capability of yielding 
sparser representations than that in \cite{BRN11}, therefore improving  
folding capacity. Access control to the folded image is realized  using 
a simple symmetric key encryption algorithm. 
The whole procedure is characterized by 
its potential for real time implementation using
parallel processing,
but also for its competitiveness using sequential processing. 
%For example: 
%the astronomical image illustrating the approach
%(size 1424$\times$ 1264$\times$ 3) has been folded 
%in encrypted form in xxx secs (in Matlab environment 
%using a small laptop)
%and recovered in xxx secs.

The paper is organized as follows. 
In  Sec.~\ref{SIR} we discuss the strategy for achieving a
high level of sparsity in image representation using   
the greedy selection strategy OMP, 
implemented here in $2$D with separable dictionaries.
The framework for extending EIF to SCEIF is discussed in 
 Sec.~\ref{SCEIF} and illustrated in
 Sec.~\ref{Expe}
by its application on  i) an astronomical image created at the
European Southern Observatory and
ii) a photograph of the natural world provided by National Geographic.
Remarks on the quality and security of the recovered images
are given in Sec.~\ref{Ana}. Conclusions and 
 final remarks and are summarized in Sec.~\ref{Con}.

\section{Sparse Image Representation}
\label{SIR}
The approach to be introduced in the next section  
relies on the ability to design a specific transformation which 
gives rise to 
a sparse representation of an image. This section is 
dedicated to the construction of such a transformation.

Suppose that an image, given as an array $I \in \R^{\Nxy}$ of intensity 
pixels, is to be approximated by the linear
decomposition
\be
I^K= \sum_{k=1}^K c_k d_{\ell_k},
\label{atom}
\ee
where each $c_k$ is a scalar and each $d_{\ell_k}$ 
is an element of $\R^{\Nxy}$ to be selected from a 
set, $\D=\{d_n\}_{n=1}^M$, 
called a `dictionary'.

A {\em{sparse  approximation}} of $I\in \R^{\Nxy}$  is 
an approximation of the form \eqref{atom} such that the number $K$
of elements in the decomposition is significantly smaller
than ${N=\Nx \Ny}$. Clearly one of the crucial issues 
to achieve high levels 
of sparsity is the  selection of the right 
elements to decompose the image. This goal has motivated 
the introduction of highly nonlinear techniques for 
image approximation, which operate outside the traditional 
basis framework. Instead, the terms in the decomposition
are taken from a large 
redundant dictionary, from where the 
elements $d_{\ell_k}$ in \eqref{atom}, called `atoms', 
are chosen according to some optimality criterion.

Within the redundant  dictionary framework for approximation,   
the problem of finding the sparsest decomposition of a given image
can be formulated as follows: 

{\em{Approximate the image  by the `atomic decomposition' 
\eqref{atom}
such that the number $K$ of atoms is minimum.}} 

Equivalently, for a dictionary of $M>N$  elements  
the statement is reworded as:

{\em{Find the atomic decomposition:
\be
I^K= \sum_{n=1}^M c_n d_n,
\label{atom2}
\ee
such that the {\em{counting measure}} $\|\vc \|_{\alpha=0}:= 
\sum_{n=1}^M (c_n)^0$ 
is minimized. }}

Unfortunately the numerical
minimization of $\|\vc \|_{\alpha=0}$  restricted to
\eqref{atom2}
involves a combinatorial problem for
exhaustive search and is therefore
intractable with classical means.
Hence, one is forced to abandon the
sparsest solution and look for a
`satisfactory solution',  i.e,
a solution such that the number of nonzero 
coefficients in \eqref{atom2} (equivalently, the number of $K$-terms 
in \eqref{atom})
is considerably smaller than the image dimension. One possibility 
for constructing a solution of this nature could be  
to fix a value of  $\alpha \in (0, 1]$  and 
minimize the diversity measure, 
$\sum_{k=1}^M |c_k|^\alpha$ \cite{REC03}, closely related
to the $\alpha$-entropy giving rise to
the non-extensive statistical mechanics
\cite{Tsa88,Tsa09}. However, the numerical implementation of this possibility is too demanding to
   apply in the present context. 
In contrast, the goal of finding a  sparse  solution 
can be achieved at speeds comparable to fast transforms by
the greedy technique called OMP
that we dedicate to be applied in $2$D. 
This approach selects the atoms in the 
decomposition \eqref{atom} in a stepwise manner, as will 
be described in the next section.
\subsection{Orthogonal Matching Pursuit in $2$D}
\label{Secomp}
OMP was introduced in 
\cite{PRK93}. We describe here our implementation in $2$D, 
henceforth referred to as OMP2D. Our version of the algorithm 
is specific to separable dictionaries, i.e, 
a $2$D dictionary which corresponds in effect to the tensor  
 product of two $1$D dictionaries. The implementation
is based on adaptive biorthogonalization and 
Gram-Schmidt orthogonalization procedures, as 
proposed in \cite{RNL02,ARN06}. 
However, the optimized selection proposed in 
\cite{RNL02} is not considered here,  
 due to the computational demands of such a selection process.

The images we are concerned with 
are assumed to be either gray level intensity images or
color images stored in a standard RGB format. 
This format stores three color values, R(Red),
G(Green) and B(Blue), for each pixel. Hence, the color image is
given as three independent $2$D arrays,
each called a `channel'. We represent the RGB channels  as
the arrays $I_{z}\in \R^{\Nxy},\, z=1,2,3$ (a
gray level intensity image 
can be considered a particular case of this representation 
corresponding to a unique index $z=1$).

Given an RGB image $I_{z} \in \R^{\Nxy},\,z=1,2,3$ and two 
$1$D dictionaries
$\D^x =\{\dx_n \in \R^{\Nx}\}_{n=1}^{\Mx}$ and
$\D^y =\{\dy_m \in \R^{\Ny}\}_{m=1}^{\My}$ our purpose is to 
approximate the arrays $I_z \in \R^{\Nxy},\,z=1,2,3$ using 
common atoms for the three images. More precisely, for $i=1,\ldots, \Nx$ and $j=i,\ldots,\Ny$
we look for approximations of the form 
\be
\label{atomml}
I_z^K(i,j)=\sum_{n=1}^{K} c_{n}^z
\dx_{\ell^x_n} (i) \dy_{\ell^y_n}(j),\quad z=1,2,3.
\ee
Notice that, while the coefficients $c_{n}^z$ in the above decomposition 
depend on the image $I_{z}$, the atoms participating 
in the decompositions are common to all the channels. For selecting those
atoms we adopt the OMP selection criterion extended to simultaneous 
decomposition of signals.  A discussion of this criterion can be found in 
\cite{TGS06}, an in our context is implemented as follows:

On setting $R_{z}^0=I_{z}, \,z=1,2,3$ 
at iteration $k+1$ the algorithm selects the atoms
$\dx_{\ell^x_{k+1}} \in \D^x$ and $\dy_{\ell^y_{k+1}} \in \D^y$ 
that maximize the sum over $z$ of the Frobenius inner 
products absolute value
$|\la \dx_n ,R_{z}^{k} \dy_m \ra_\F|,\, n=1,\ldots,\Mx,\,m=1,\ldots,\My, 
$ i.e.,
\be
\begin{split}
{\ell^x_{k+1},\ell^y_{k+1}}&= \operatorname*{arg\,max}_{\substack{n=1,\ldots,\Mx\\
m=1,\ldots,\My}} \sum_{z=1}^3|\sum_{\substack{i=1\\j=1}}^{{\Nx,\Ny}} 
\dx_{n}(i) R_{z}^k(i,j)\dy_{m}(j)|,\\
\text{with}\\
R_{z}^{k}(i,j)& = I_{z}(i,j) - \sum_{n=1}^{k} c^{z}_{n}
\dx_{\ell^x_n} (i) \dy_{\ell^y_n}(j),\quad z=1,2,3.
\end{split}
\label{ompml}
\ee
The three sets of coefficients $c_{n}^{z},\,n=1,\ldots,k$ involved in 
\eqref{ompml}
are such that 
$\|R_{z}^{k}\|_\F$ is minimum for each $z$. 
($\|\cdot \|_\F$  being the Frobenius norm). 
This is guaranteed by 
calculating the coefficients $c_n^z,\, z=1,2,3$ 
as 
\be
\label{Bcoe}
c_n^z= \la B_n^k, I_{z} \ra_\F,\quad n=1,\ldots,k,
\ee
where matrices $B_n^k,\,n=1,\ldots,k,$ are recursively constructed at 
each iteration step as indicated in Appendix A. 

The algorithm iterates up to step, say $K$, for which, for a given $\rho$,
the stopping criterion $\sum_{z=1}^3||I_{z} - I_{z}^K||^2_\F < \rho$ is met.
The MATLAB function for the implementation of the
OMP2D approach on multiple $2$D signals, which we have called OMP2DMl, 
is available from \cite{Webpage2}. 
The corresponding MEX file in C++, for faster implementation of the identical function, is also
available from \cite{Webpage2}.
%Other Matlab functions and MEX files 
%for more sophisticated Pursuit Strategies refining the search 
%\cite{RNL02, ARN05,ARN06} are also available at \cite{Webpage2}.
%All such refinements are computationally more 
%demanding, though. We have found OMP2D to be the most suitable selection 
%technique for the dictionaries we are considering in  this
%particular application.
\subsection{Constructing the dictionary}
\label{Secdict}
The other crucial design for success in finding a
`good enough'  sparse representation of the form \eqref{atomml} 
is the dictionary  which provides the 
possible choices of atoms at the selection step.

\begin{figure}[!h]
%\begin{figure}[!h]
\begin{center}
\includegraphics[width=8cm,height=5cm]{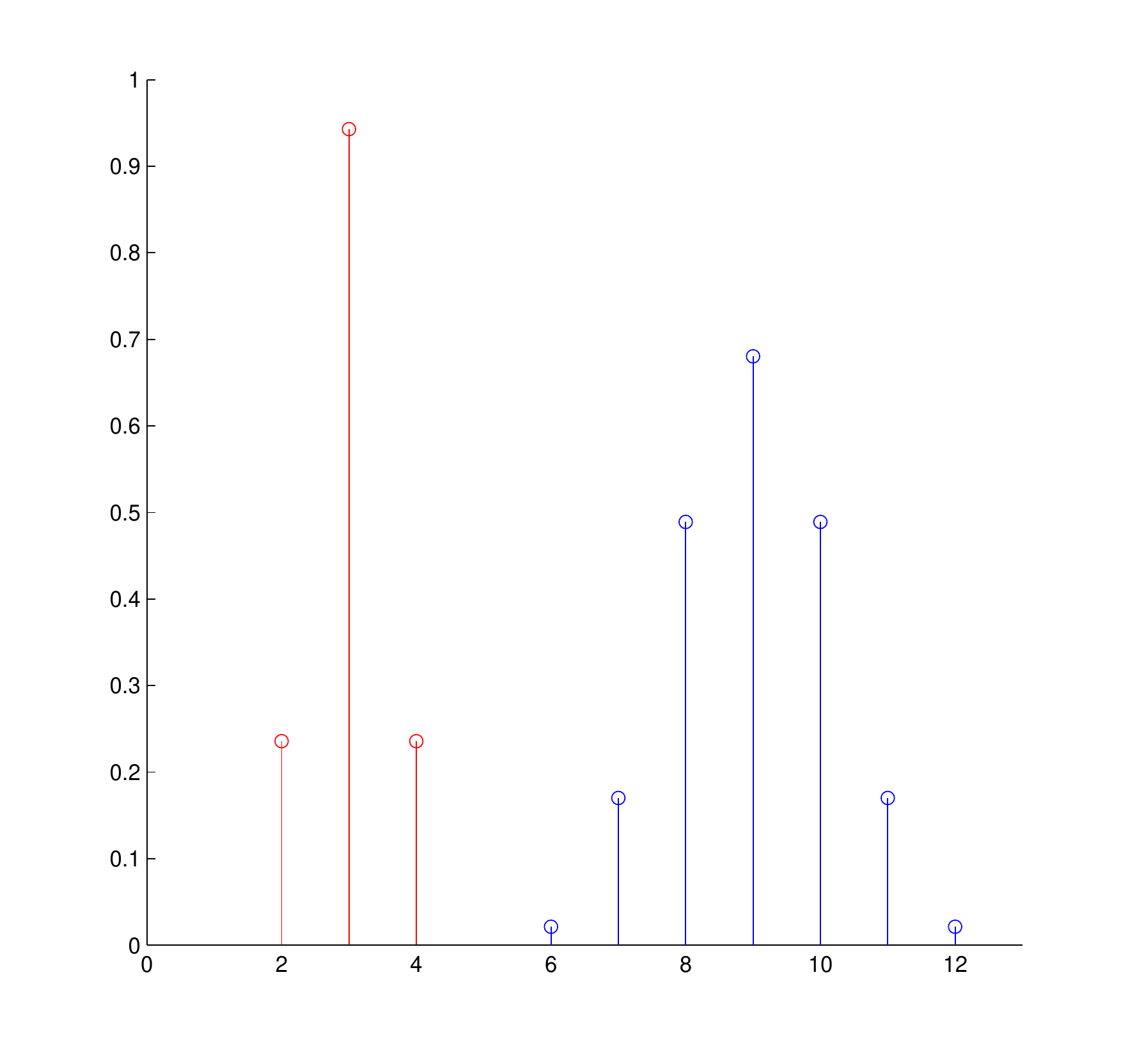}
\end{center}
\vspace{-0.75cm}
\caption{\small{The discrete prototype cubic B-splines of 
supports 3, and 7, generating (by translation at every 
 point) the dictionaries $\Dx_3$ and $\Dx_4$, respectively.}}
\label{splinesa}
\end{figure}

The mixed dictionary used in \cite{BRN11} for this purpose 
consists of two components for 
each $1$D dictionary:
\begin{itemize}
\item
A Redundant Discrete Cosine dictionary (RDC) 
$\D^x_1$ as given by:
$$\D^x_1=\{w^c_i\cos(\frac{\pi(2j-1)(i-1)}{2\Mx}),\,j=1,\ldots,\Nx
\}_{i=1}^{\Mx},$$
with $w^c_i,\,i=1,\ldots,M_x$ normalization factors.
For $\Mx=\Nx$ this set is a Discrete Cosine orthonormal
basis for the Euclidean space $\R^{\Nx}$.
For $\Mx=2l\Nx$, with $l \in \N$, the set is an RDC dictionary 
with redundancy $2l$, that will be fixed equal to 2.
\item
The standard Euclidean basis, also called the Dirac basis, i.e. 
$$\Dx_2=\{e_i(j)=\delta_{i,j},\,j=1,\ldots,\Nx\}_{i=1}^{\Nx}.$$
\end{itemize}
Now we include an additional component:
\begin{itemize}
\item
A family of cubic B-spline 
dictionaries of different support, as proposed in \cite{ARN05}, but 
discretizing the domain by taking the value of a
prototype  B-spline only at the knots 
and translating that
prototype one point at each translation step.
Each B-spline based dictionary is given as
$$\Dx_s=\{w^s_iB^s_m(j-i)|\Nx; j=1,\ldots,\Nx\}_{i=1}^{\Mx^s},$$
where the notation $B^s_m(j-i)|\Nx$ indicates the restriction of the 
B-spline of order $m$, centered at the point $i$, to be
an array of size $\Nx$. Cubic splines are obtained setting 
$m=4$.
The factors $w^s_i,\,i=1,\ldots,{\Mx}^s$ are normalization constants,  
with $\Mx^s$ the number of atoms in the dictionary $s$.
The values of $s$ to be considered are $s=3$ and $s=4$, which 
label the dictionaries arising as translation of a prototype 
B-spline having, respectively, 3 and 7 points of nonzero value 
(see Fig.~\ref{splinesa}).
\end{itemize}
The complete $1$D dictionary is constructed as
$\Dx= \cup_{s=1}^4 \Dx_s$.
The dictionary $\Dy$ is  built in  equivalent fashion, but
changing $\Nx$ to $\Ny$ and $\Mx$ to $\My$ when applicable.

The required $2$D dictionary is formed as 
$\D=\D^x \otimes \D^y$. However, it is not necessary to store the $2$D 
dictionary $\D$, since the algorithm takes advantage of
the separability inherent in its construction.
This advantage significantly reduces storage demands 
and extends the possibility of using the OMP approach in $2$D.

It is time now  to examine closely the term
 `good enough' for 
a sparse decomposition. Within the present context by  
 the term good enough we mean a decomposition that 
a)increases sparsity well beyond the levels attained by such 
techniques as  DCT or 
DWT, and b)requires 
comparable computational time.
\begin{remark}
The suitability of the mixed dictionary for block processing is essential in fulfilling requirements
a) and b) above, i.e., for processing an image by dividing 
it into {\em{small blocks}} and approximate the blocks independently.
This feature renders the complexity of the highly nonlinear,
and otherwise costly selection technique, 
{\em{linear}} in terms of the number of blocks 
employed in decomposing the image.
\end{remark}
The capacity of the dictionary based approach to achieve a satisfactory 
sparse approximation of an image will become 
clear when illustrating the SCEIF technique in Sec.~\ref{Expe}.
In addition, we present some comparisons on the results on 
standard test images
which are listed in the first column of
Table \ref{table:cr}.
All the images are 8-bit gray level intensity images of
$512 \times 512$ pixels.
For the actual processing we divide each  
image into blocks of $8 \times 8$ pixels and process the blocks 
independently. The approximated blocks are then assembled to give the 
approximated image.
\begin{table}
\begin{center}
\begin{tabular}{ | l || c | c | c | c | c | }
\hline
Image & Dictionary & DCT & DWT\\ \hline \hline
Barbara & 5.02 & 3.10 & 2.94   \\ \hline
Boat & 4.61 & 2.61 & 2.60 \\ \hline
Bridge & 3.24 & 1.79 & 1.86 \\ \hline
Film Clip & 5.86 & 3.29 & 3.34 \\ \hline
Lena & 6.51 & 3.81 & 4.04 \\ \hline
Mandrill & 2.85 & 1.64 & 1.64 \\ \hline
Peppers  & 5.23 & 2.88 & 2.96 \\ \hline
\end{tabular}
%\vspace{-0.55cm}
\caption{\small{Comparison of the Sparsity Ratio (for PSNR $43$dB )
achieved by the mixed dictionary (second column) and 
that yielded by DCT and DWT (3rd and 4th columns 
respectively). The first column lists the names of the 
popular test images where the approaches are compared.}}
\label{table:cr}
\end{center}
\end{table}
Sparsity is measured by 
the Sparsity Ratio (SR) defined as
$${\text{SR}}=\frac{\text{total number of pixels}}{\text{total number of
coefficients}}.$$
In all the cases the number of coefficients is determined
as the one required to produce a high quality approximation with no
visual deterioration with respect to the original image,
in this case corresponding to a PSNR of $43$dB (c.f. \eqref{psnr}).
The sparsity results achieved by selecting atoms with OMP2D, from the 
proposed  mixed dictionary, are displayed in  
the second column of Table \ref{table:cr}.
The third column shows
results produced by the DCT implemented using the same blocking
scheme. For further comparison the results produced by
the Cohen-Daubechies-Feauveau 9/7 DWT
(applied on the whole image at once) are displayed in the
last column of Table \ref{table:cr}. Notice that 
while for the fixed
PSNR  of $43$dB the DCT and DWT approaches yield comparable SR,
the corresponding SR obtained by the mixed dictionary,
for all the images, is significantly higher.
What is of paramount importance to our current interest is that the 
processing time is very competitive. 
The actual speed of the approximation  depends, of course, 
on the sparsity of each image. For the set of images in Table \ref{table:cr}
the mean SR is 4.76 and the mean processing time is
1.72 seconds per image (average of ten independent runs 
in MATLAB environment implemented in a 14'' laptop with a 
 2.8 GHz processor and 3GB RAM).
\section{Self Contained Encrypted Image Folding}
\label{SCEIF}
The idea of using the null space of a
transformation for storing information in encrypted form was
first outlined in \cite{MRN04} and further discussed in
\cite{RNP06}. However, it has been only recently materialized as
the EIF application \cite{BRN11}.
The denomination is meant to reflect a particular feature;
the space created by a sparse representation of an image is
used to store part of the image itself,
thereby reducing the original image size.

As already stated, we process each image $I_z,\,z=1,2,3$ 
by dividing it into, say $Q$, blocks $I_{z,q},\,q=1,\ldots,Q$,  which  
without loss of generality are assumed to be square of
$N_q\times N_q$ intensity pixels.  For a fixed $q$-value 
the three blocks of intensity arrays $I_{z,q},\, z=1,2,3$
(each of which corresponds to a color channel)
are simultaneously approximated using the 
dictionary $\D= {\Dx} \otimes {\Dy}$, as given in  Sec~\ref{Secdict},
by the atomic decomposition
\be
\label{repbq}
I_{z,q}^{K_q}=\sum_{n=1}^{K_q} c_{n}^{z,q}
\dx_{{{\ll^x_n}^q}} \dy_{{{\ll^y_n}}^q},
\quad q=1,\ldots,Q,\, z=1,2,3
\ee
where $\dx_{{\ll^x_n}^q}$ and $\dy_{{\ll^y_n}^q},\,n=1,\ldots,K_q$ 
are the atoms that 
have been selected  through the approach of Sec~\ref{Secomp} 
and span a subspace 
$\V_{K_q}=\Spann\{ \dx_{{\ll^x_n}^q} \otimes \dy_{{\ll^y_n}^q}\}_{n=1}^{K_q} 
\subset \R^{N_q\times N_q}$. 
%The notation
%$c^{K_q}_n,\,n=1,\ldots, K_q$ is used to indicate the coefficients
%of the atomic decomposition in the $q$th-block.

For \eqref{repbq} to be a sparse approximation of $I_{z,q}$ the number 
of $K_q$ terms should be considerably smaller 
than $N_q^2$. In other words, the dimension 
$N_q^2- K_q$ of the orthogonal complement 
of $\V_{K_q}$ in $\R^{N_q\times N_q}$, which is indicated as
$\V_{K_q}^\bot$, should be significant in relation to 
$N_q^2$. In line with \cite{BRN11} the subspace 
$\V_{K_q}^\bot$ is used to embed a part of the image in another
part of the image, as described below. The approximated 
image $I_z^K= \cup_{q=1}^Q I_{z,q}^{K_q}\, z=1,2,3$ 
is the {\em{plain text}} and the {\em{cipher}} is the folded image.

\subsection{Folding Procedure}
A number of, say $3H$, blocks are kept 
as `hosts' for embedding
the coefficients of the remaining $3(Q-H)$ equations \eqref{repbq}. 
For this, 
first the coefficients $c_{n}^{z,q}, n=1,\ldots, K_q, \,q= 
(H+1),\ldots, Q,\,z=1,2,3$ are relabeled to became the components 
of vectors $(h_1^{z,q},\ldots,h_{L_q}^{z,q}), q= 1,\ldots,H,\,z=1,2,3$, 
each of length $L_q=N_q^2-K_q$.
These vectors are embedded  in the 
$3H$ host blocks, according to the procedure 
given in \cite{BRN11}, as follows.
\begin{itemize}
\item
For each value of $q$ and $z$  build  a block
of pixels $F_{z,q} \in \R^{N_q \times N_q}$ as
\be
\label{plain}
F_{z,q}=\sum_{i=1}^{L_q}  h_i^{z,q} U_i^{z,q},\quad q= 1,\ldots,H,\,z=1,2,3
\ee
where $U_i^{z,q} \in \R^{N_q \times N_q},\,i=1,\ldots,L_q$  is 
an orthonormal basis for $\V_{K_q}^\bot$ obtained as follows:
\begin{itemize}
\item[a)]Using matrices $Y_i^{z,q} \in \R^{N_q\times N_q},\,i=1,\ldots,L_q$
randomly generated, with a {{\em public}} initialization ${\seed}$, 
and the already constructed projector $\op_{\V_{K_q}}$ 
(c.f.\eqref{GS}), for $q=1,\ldots,H$ and $z=1,2,3$
compute the matrices $O_i^{z,q}$ as 
\be
\label{Ymat}
O_i^{z,q} = Y_i^{z,q} - \op_{\V_{K_q}} Y_i^{z,q} 
\in \V_K^\bot,\quad i=1,\ldots,L_q. 
\ee
\item[b)]
Transform these matrices, using a random  transformation
${\hat{\Pi}}_{{\key}}$ initialized with 
a {\em{private}} $\key$, to obtain a private set of matrices
\be
\label{Xmat}
{\hat{\Pi}}_{{\key}}:(O_i^{z,q},\, i=1,\ldots,L_q) \rightarrow \{X_i^{z,q}\}_{i=1}^{L_q}.
%X_i^{z,q}= 
%{\hat{\Pi}}_{{\key}}O_i^{z,q},\quad i=1,\ldots,L_q.
\ee
\item[b)]
For each $z$ and $q$ 
use an orthonormalization procedure, that we indicate by the 
operator $\orth(\cdot)$, to
orthonormalize matrices $X_i^{z,q},\,
i=1,\ldots,L_q,$ and have the orthonormal basis 
%\begin{comment}
%\be
%\label{Umat}
%U_i^{z,q}= {\orth}(X_i^{z,q},\,i=1,\ldots,L_q),\quad q=1,\ldots,H,\,
%z=1,2,3
%\ee
%$$\{U_n^{z,q} \}_{n=1}^{L_q} = {\orth}\{X_i^{z,q},\,i=1,\ldots,L_q\},\quad q=1,\ldots,H,\,z=1,2,3$$
%$$\{U_n^{zq}, \, n=1,\ldots, Lq \} = {\orth}\{X_i^{z,q},\,i=1,\ldots,L_q\},\quad q=1,\ldots,H,\,z=1,2,3$$
%$$\{U_n^{zq} \}_{n=1}^{L_q}= {\orth} \{X_i^{zq} \}_{i=1}^{L_q}$$
%\end{comment}
\be
\label{Umat}
\{U_i^{z,q}\}_{i=1}^{L_q} = {\orth}(X_i^{z,q},\,i=1,\ldots,L_q),\quad q=1,\ldots,H,\,z=1,2,3
\ee
to be used in \eqref{plain}
for embedding the coefficients of the remaining blocks
$I_{z,q}^{K_{q}},\, q=(H+1),\ldots,Q,\,z=1,2,3$.
\end{itemize}
%\item[b)]Setting an initialization ${\key_2}$, which remains
%{\em{unknown}} to unauthorized users, apply
%a random rotation $\Pi_{{\key_2}}$ on matrices
%$O_i^{z,q}$ to generate the orthonormal basis
%\be
%\label{Umat}
%U_i^{z,q}= \Pi_{{\key_2}} O_i^{z,q},\,i=1,\ldots,L_q,\, q=1,\ldots,H,\, 
%z=1,2,3
%\ee
%to be used in \eqref{plain}
%for embedding the coefficients of the remaining blocks
%$I_{z,q}^{K_{q}},\, q=(H+1),\ldots,Q,\,z=1,2,3$.
%\end{itemize}
\item
Fold the image by the superpositions
$G_{z,q}= I_{z,q}^{K_q} +F_{z,q}, q=1,\ldots,H, \, z=1,2,3$ 
and subsequent composition
$G_z = \cup_{q=1}^H G_{z,q},\, z=1,2,3$.
\end{itemize}
%Notice that the folded image $G$ is an element of $\R^{HN_q \times HN_q}$ 
%and $HN_q \times HN_q$ is smaller than the size
%$Q N_q \times Q N_q$ of the original images $I$. How much 
%smaller  depends on  the sparsity of the
%representations \eqref{repbq}. 
%The sparser the representations  the smaller the size of the folded image 
%in relation to the original one.
{\subsubsection{Making the approach self contained}}
Knowledge of the coefficients in 
\eqref{repbq} is not enough to reconstruct the blocks 
$I_{z,q}^{K_q},\, q=1,\ldots,Q ,\,z=1,2,3$. 
For each $q$-value it is also necessary to 
know the indices of the atoms in the decomposition. 
This matter is not considered in \cite{BRN11}. A  
contribution of this effort is the generalization of the previous 
approach to deal with the storage of indices as well. 
The present proposal consists of creating some `ad hoc' blocks
to embed the required indices. 
Without loss of generality the blocks are assumed to
be square containing $\Nt_q \times \Nt_q$ intensity pixels.
Using {\em{any}} atom normalized to unity, 
say $A_q \in \R^{\Nt_q \times \Nt_q}$, the
ad hoc intensity arrays $\It_q \in
\R^{\Nt_q \times \Nt_q},\,q=1,\ldots \Qt$ are 
created  as
\be
\label{adhoc}
\It_q= K_q A_q,\quad q=1,\ldots,\Qt,
\ee
and $\Lt_q=\Nt_q^2-1$ indices are embedded in the orthogonal complement 
(with respect to $\R^{\Nt_q \times \Nt_q}$) of the subspace 
spanned by the single atom $A_q$.
The embedding procedure is equivalent 
to that for embedding the coefficients, i.e.,
\begin{itemize} 
\item
For $q=1,\ldots,\Qt$ using a {{\em public}} 
initialization ${\seed}$ generate the random matrices
$\Yt_i,\,i=1,\ldots,\Lt_q$ to calculate the matrices 
$\tilde{O}_i^q$ as
\be
\label{Ytmat}
\tilde{O}_i^q = {\Yt_i^q -   A_q \la A_q, \Yt_i^q \ra_\F} ,\quad i=1,\ldots,\Lt_q.
\ee
\item
Transform these matrices, using a random transformation initialized with
the {\em{private}} $\key$, to obtain a private set of matrices
\be
\label{Xtmat}
{\hat{\Pi}}_{{\key}}:(\tilde{O}_i^q,
\,i=1,\ldots,\tilde{L}_q) \rightarrow \{\tilde{X}_i^q\}_{i=1}^{\tilde{L}_q}.
%\Xt_i^{q}=
%\Pi_{{\key}}\tilde{O}_i^q,
%\quad i=1,\ldots,\Lt_q.
\ee
\item
For each $q$-value use the orthonormalization procedure  
$\orth(\cdot)$ to orthonormalize matrices $\Xt_i^{q},\,
i=1,\ldots,\Lt_q,$  to have the orthonormal basis
\be
\label{Umatt}
\{\Ut^{q}_i\}_{i=1}^{\Lt_q} = \orth(\Xt_i^{q},\,i=1,\ldots,\Lt_q), \quad q=1,\ldots,\Qt,
\ee
needed to embed the indices. For this, 
first map each ordered pair of indices 
$(n,m),\, n=1,\ldots, \Mx^q,\, m=1,\ldots, \My^q$ (which label the
$2$D dictionary atoms) to the single label 
$\til{n}=1,\ldots, \Mx^q \My^q$. Now the steps for 
embedding the indices of the atoms  
in $I_{z,q}^{K_q},\, q=1,\ldots,Q$ (c.f. \eqref{repbq}) 
parallel those for embedding the coefficients.    
Arrange the indices
to be components of vectors 
$(\htt^q_1,\ldots,\htt^q_{\Lt_q}),\,q=1,\ldots,\Qt$. For each 
$q$-value, use the corresponding vector to 
generate the block of pixels 
$\Ft_q \in \R^{\Nt_q \times \Nt_q}$ as 
\be
\label{plaint}
\Ft_q=\sum_{i=1}^{\Lt_q} \htt^q_i \Ut^q_i,\quad q=1,\ldots, \Qt.
\ee
\item
Now `fold' the  ad hoc  blocks by the superpositions
$\Gt_q= \It_q +\Ft_q, q=1,\ldots,\Qt$ and  subsequently 
produce the composition
$\Gt = \cup_{q=1}^{\Qt} \Gt_q$  to be 
split into three  channels $\Gt_z,\,z=1,2,3$.
\end{itemize}
The folding process finishes by joining the folded channels 
$G_z,\,z=1,2,3$ and the ad hoc ones $\Gt_z,\,z=1,2,3$ 
to create the single folded RGB image $I_{\rm{folded}_z},\,z=1,2,3$ as 
$$I_{\rm{folded}_z}= G_z \cup \Gt_z,\, z=1,2,3.$$
This image is now endowed with all the information that is needed to 
recover the approximation of the original image. 

Note: Parameters, such as the public $\seed$ and the original
image dimensions which would normally be placed in the
header, are added as pixel values in the last row of the folded image.

\subsection{Recovering Procedure}
\label{RP}
At this stage the approximation $I_z^K= \cup_{q=1}^{Q} I_{z,q}^{K_q},
\,z=1,2,3$ 
of the RGB image $I_z,\,z=1,2,3$ is recovered from the 
folded  RGB image $I_{\rm{folded}_z},\,z=1,2,3$ by following the 
steps below.
\begin{itemize}
%\item
%Obtain, from the last intensity pixels of the folded image 
%the parameters, i.e.
%number of parameters, $xx1\, xx2,\, etc$. 
\item 
Separate $I_{\rm{folded}_z}$ into $G_z,\,z=1,2,3$ and $\Gt$, and these into 
the blocks $G_{z,q},\,q=1,\ldots,H,\,z=1,2,3$ 
and $\Gt_q,\,q=1,\ldots,\Qt$. 
\item
Obtain $K_q,\,q=1,\ldots,\Qt$ from the 
inner products $\la A_q,\Gt_q\ra_\F= K_q,\,q=1,\ldots,\Qt$ (the 
remaining ones, $K_q,\,q=\Qt+1,\ldots,Q$, can be hidden 
in some additional ad hoc blocks or just given as plain text 
intensity pixels).
\item
Obtain
$\Ft_q$ as $ \Ft_q= \Gt_q -  K_q A_q,\, q=1,\ldots,\Qt$.
\item
Recover the indices $(\htt^q_1, \ldots, \htt^q_{\Lt_q}),
q=1,\ldots,\Qt$ as
$$\htt^q_i = \la \Ut^q_i, \Ft_q \ra_\F, \quad i=1,\ldots, {\Lt_q},$$
and  map them back to the 
arrays of ordered pairs  $\{({\ll^x_n}^q, {\ll^y_n}^q)\}_{n=1}^{K_q},\, q=1,\ldots,Q$.
\item
Obtain $I_{z,q}^{K_q},\, q=1,\ldots,H,\,z=1,2,3$ from $G_{z,q}$ as
$I_{z,q}^{K_q}= \op_{\V_{K_q}} G_{z,q}$ and $F_{z,q}$ as
$F_{z,q}= G_{z,q}-I_{z,q}^{K_q},  \,q=1,\ldots,H,\,z=1,2,3$.
\item
Recover vectors $(h^{z,q}_1, \ldots, h^{z,q}_{L_q}),\, q=1,\ldots,H,\,
z=1,2,3$
as 
$$h^{z,q}_i = \la U^{z,q}_i, F_{z,q} \ra_\F, \quad i=1,\ldots, {L_q},$$
and regroup them back to get the original arrays of coefficients $\{c^{z,q}_{n}\}_{n=1}^{K_q},\, 
q=(H+1),\ldots,Q,\,z=1,2,3.$
\item
Use the recovered indexes and the recovered    
coefficients to compute $I_{z,q}^{K_q},\,q=(H+1),\ldots,Q,\,z=1,2,3$ 
as in \eqref{repbq} and reconstruct the 
approximated RGB image $I_z^K$ as
$$I_z^K=\cup_{q=1}^Q I_{z,q}^{K_q},\,z=1,2,3.$$
\end{itemize}
%\begin{remark}
%Notice that within the proposed framework the order (labels) of 
%the embedded coefficients,
%indices, and number of atoms per block can be arbitrary. This
%leaves room for further privacy. We use the 
%possibility by ordering the corresponding labels using a 
%random permutation initialized by the secret key. However, 
%in the current implementation of the approach 
%the scrambling of these parameters is not crucial as far as security is 
%concerned. The discussion of security issues is postponed to 
%section Sec.\ref{security}.
%\end{remark}
\section{Numerical Examples}
\label{Expe}

\begin{figure}
  \begin{center}
    \includegraphics[width=0.32\textwidth]{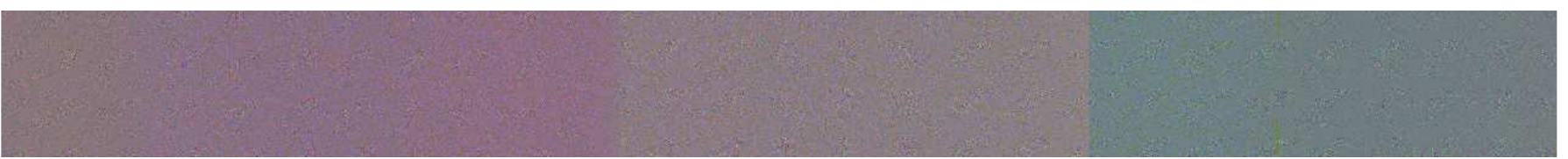}

    \includegraphics[width=0.32\textwidth]{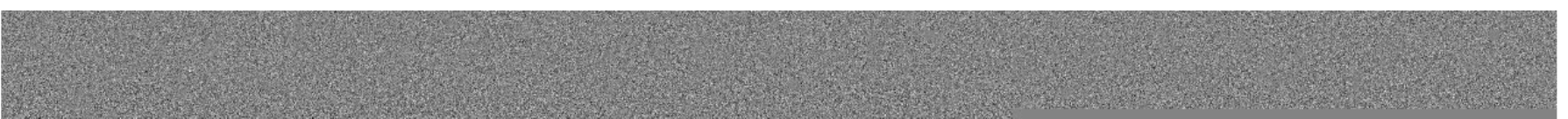}

    \includegraphics[width=0.32\textwidth]{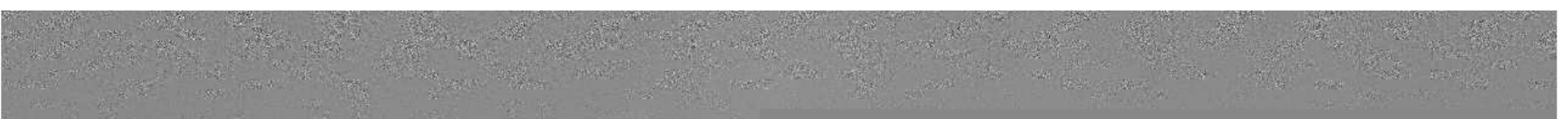}
    \includegraphics[width=0.32\textwidth]{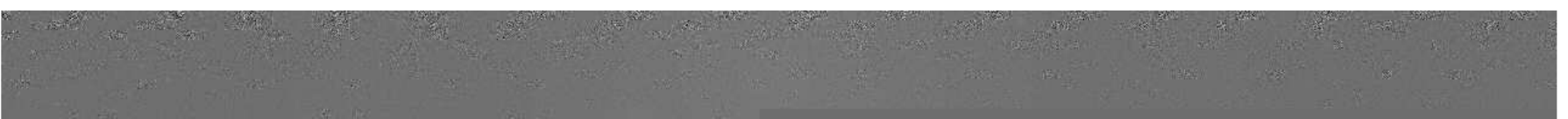}
    \includegraphics[width=0.32\textwidth]{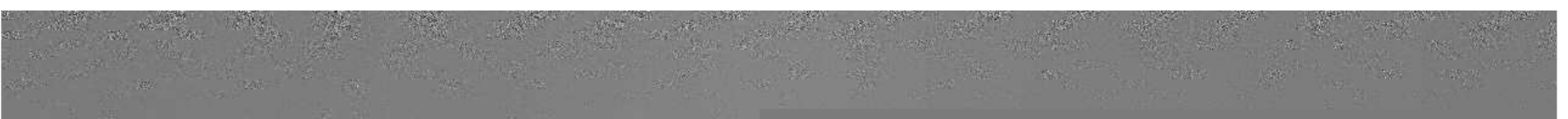}

    \includegraphics[width=0.32\textwidth]{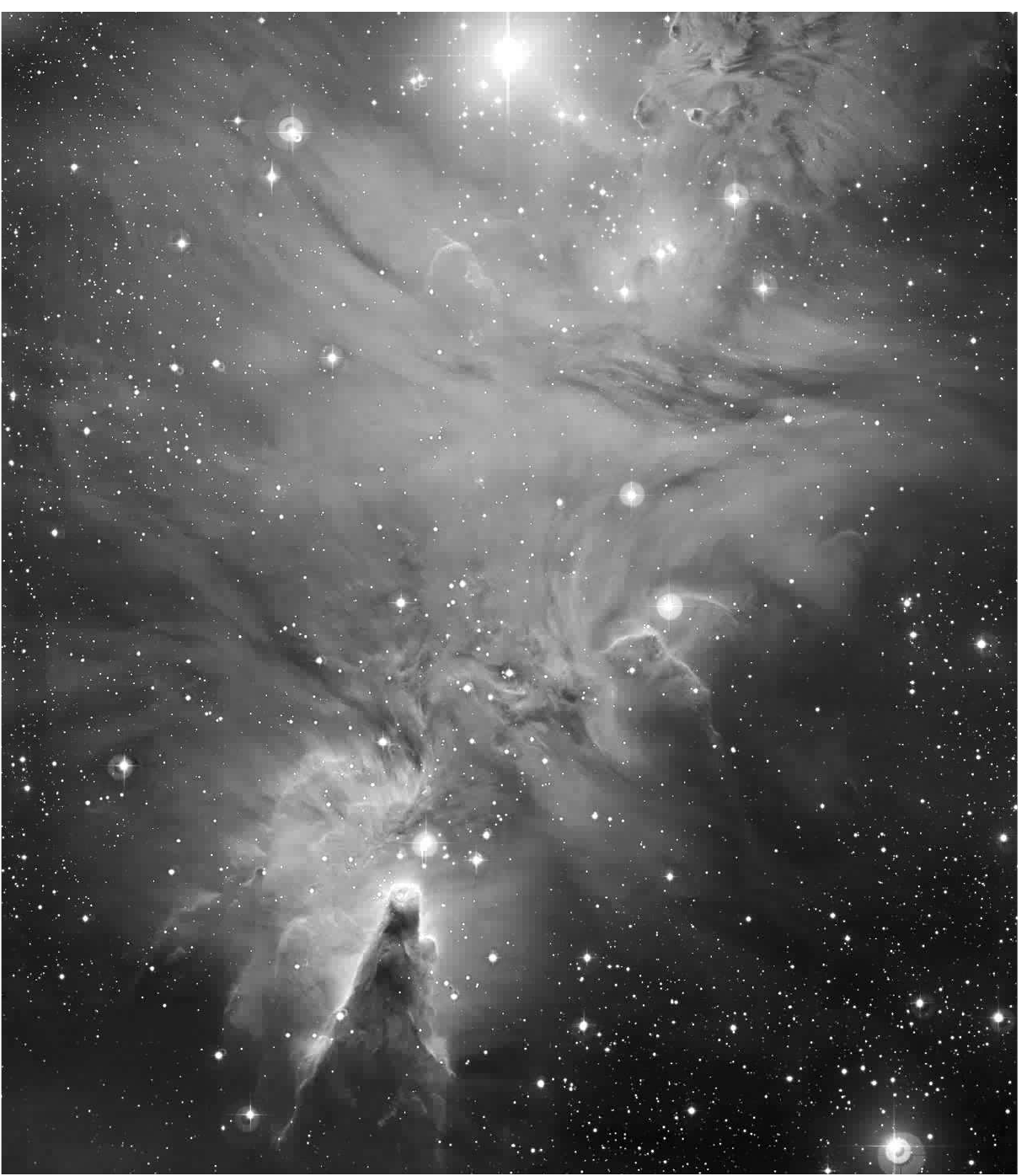}
    \includegraphics[width=0.32\textwidth]{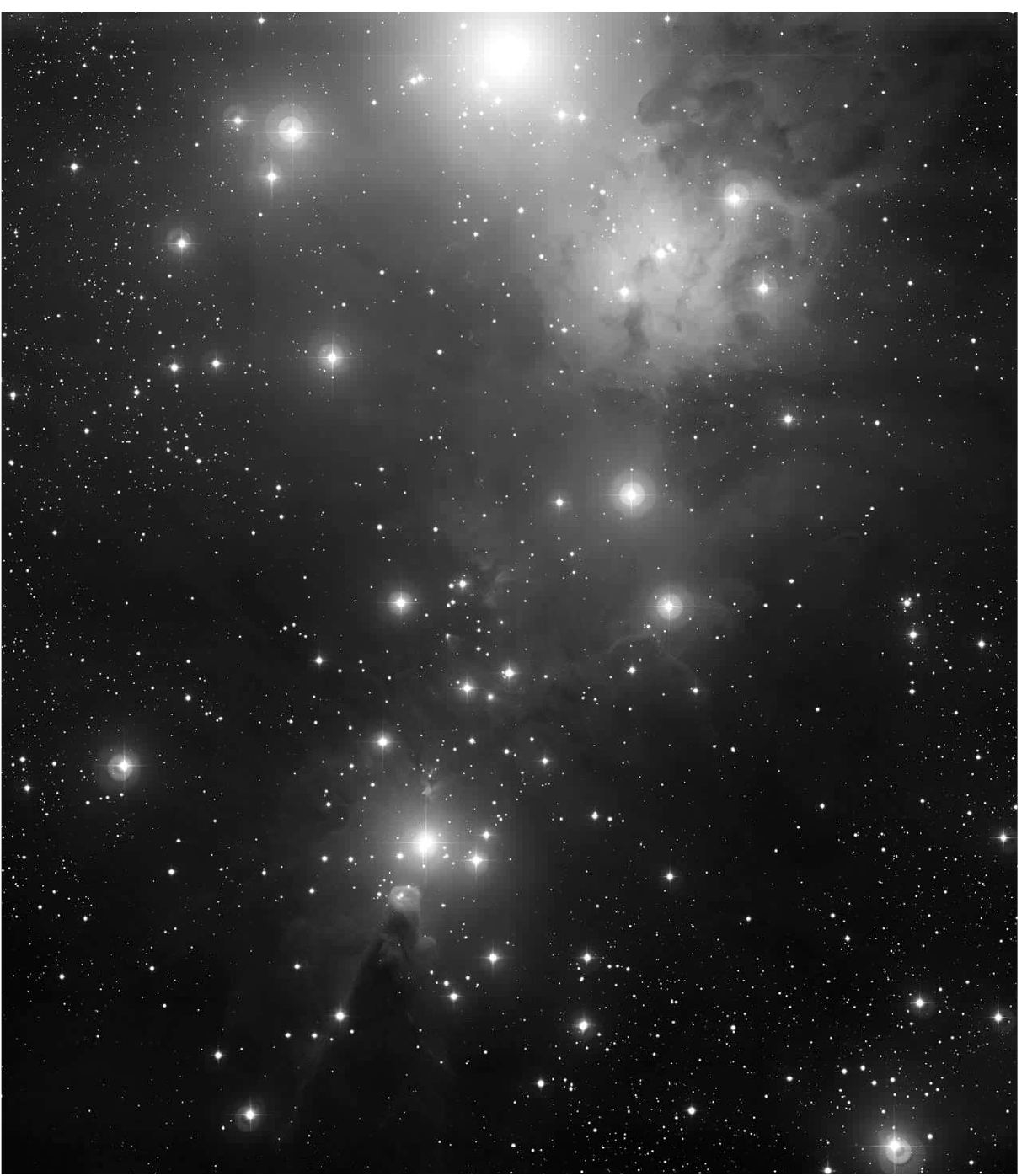}
    \includegraphics[width=0.32\textwidth]{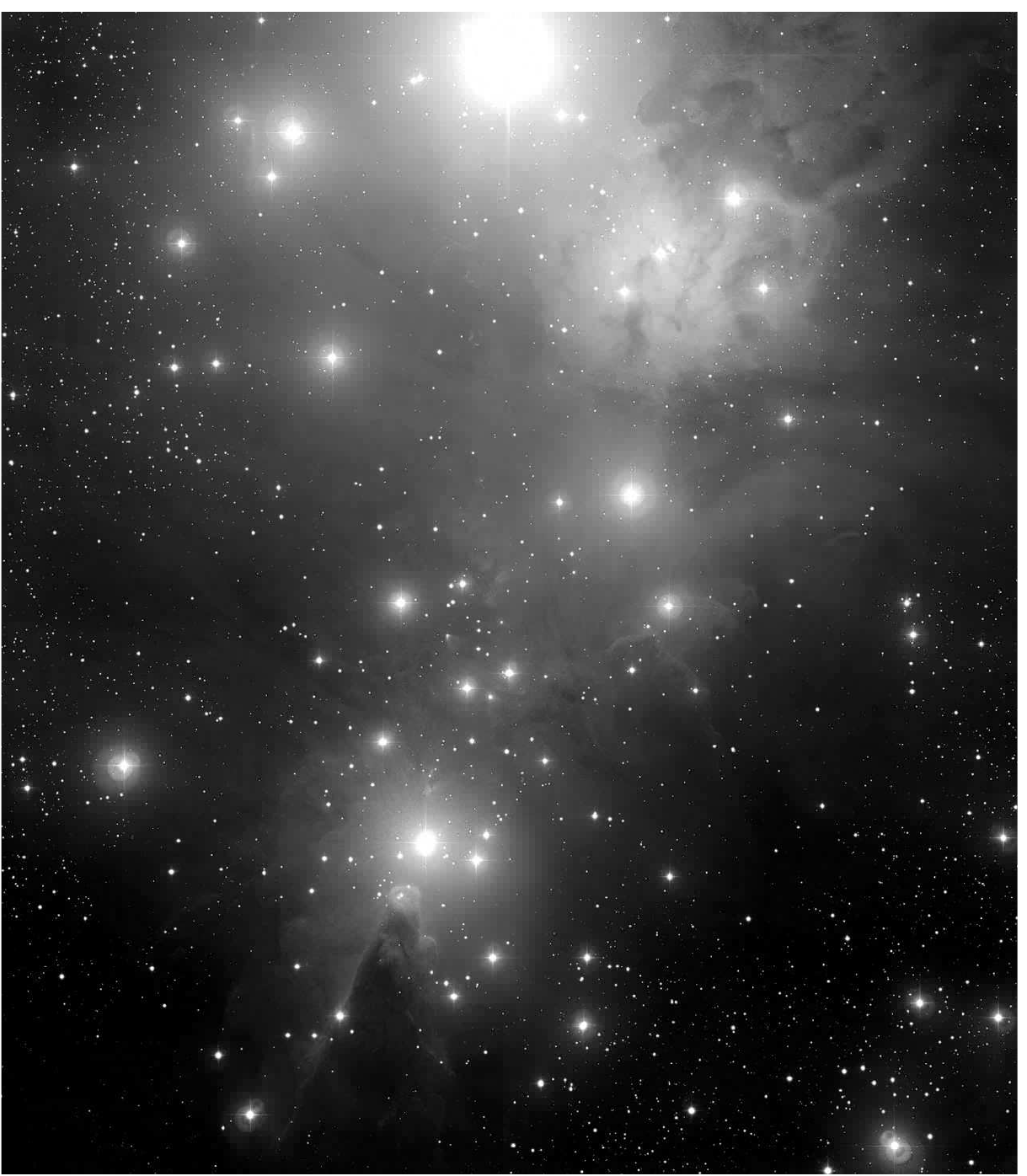}

    \includegraphics[width=0.32\textwidth]{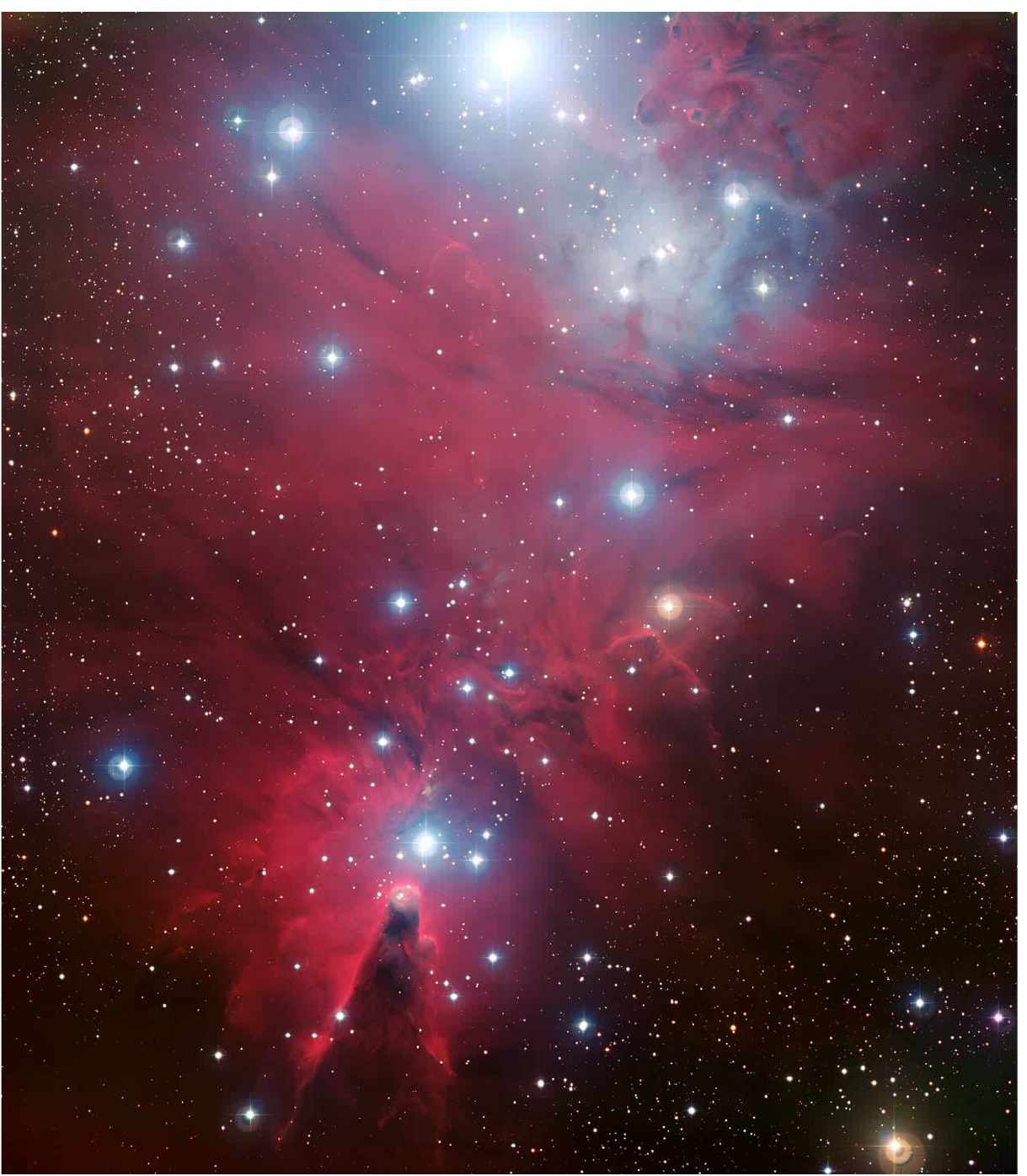}
  \end{center}
 \caption{\small{The  color image represents a high quality approximation 
 (PSNR 42.5 dB) of an image of the nebula
 NGC 2264.  Credit ESO \cite{ESOWeb}}.  The small picture at the top is the RGB folded image. The 
 one right below is the part containing the indices.
The three small pictures in the next row are the folded channels  
(each of which contains coefficients of 
plain text representation of that channel).
The three larger pictures  are the channels recovered from the previous
ones. The bottom picture is the recovered RGB image. The 
 recovery is successful because it was realized with the authorized 
 key.}
\end{figure}
In this section the SCEIF approach is illustrated with two examples both
involving  an RGB color image.

The picture at the bottom of Fig.~2 is an image of the nebula  
NGC 2264 created at the European Southern Observatory 
(ESO) \cite{ESOWeb}. 
The resolution of this image is $1464 \times 1280$ pixels per channel.
The $2$D intensity arrays, one for each channel, 
are the three pictures  right above the color 
one.
In order to apply SCEIF firstly each channel is divided 
into small blocks of $8 \times 8$ pixels. The blocks are 
approximated using the mixed dictionary of Sec.\ref{Secdict}
and the approach of Sec. \ref{Secomp}.
The approximation is of high quality. This is ensured by 
using two measures on the whole color image: 
a high PSNR (42.5 dB) and a high 
Mean Structural Similarity Index
(0.997) \cite{ssim} (further comments are given in 
Sec.~\ref{qual}). Each channel in Fig.~2 is 
folded and reshaped to produce a single 
 RGB image. The latter is the small 
picture at the top of Fig.~2. Notice 
that the size of such an image is `extra small' 
($120 \times 1280  \times 3$ pixels) in comparison to 
the original ($1464 \times 1280 \times 3$ pixels).
This is because the representation of the full image 
by the proposed mixed dictionary is very sparse. 
The SR  for the image is 17.42.
\begin{figure}
  \begin{center}
    \includegraphics[width=0.32\textwidth]{SCEIF_Images_Astro/astro_dict_folded}

    \includegraphics[width=0.32\textwidth]{SCEIF_Images_Astro/astro_dict_index_folded}

    \includegraphics[width=0.32\textwidth]{SCEIF_Images_Astro/astro_dict_folded_R}
    \includegraphics[width=0.32\textwidth]{SCEIF_Images_Astro/astro_dict_folded_G}
    \includegraphics[width=0.32\textwidth]{SCEIF_Images_Astro/astro_dict_folded_B}

    \includegraphics[width=0.32\textwidth]{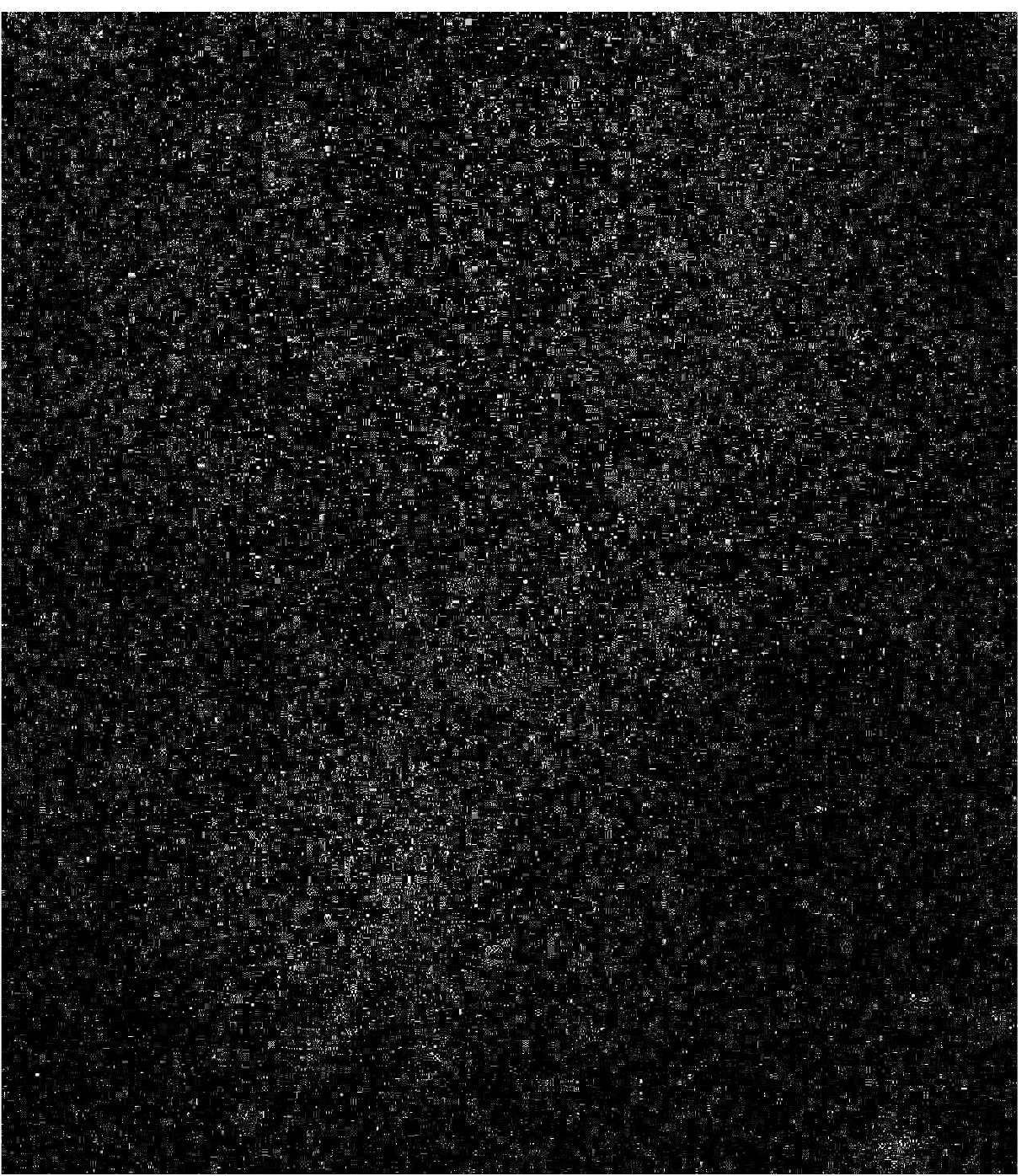}
    \includegraphics[width=0.32\textwidth]{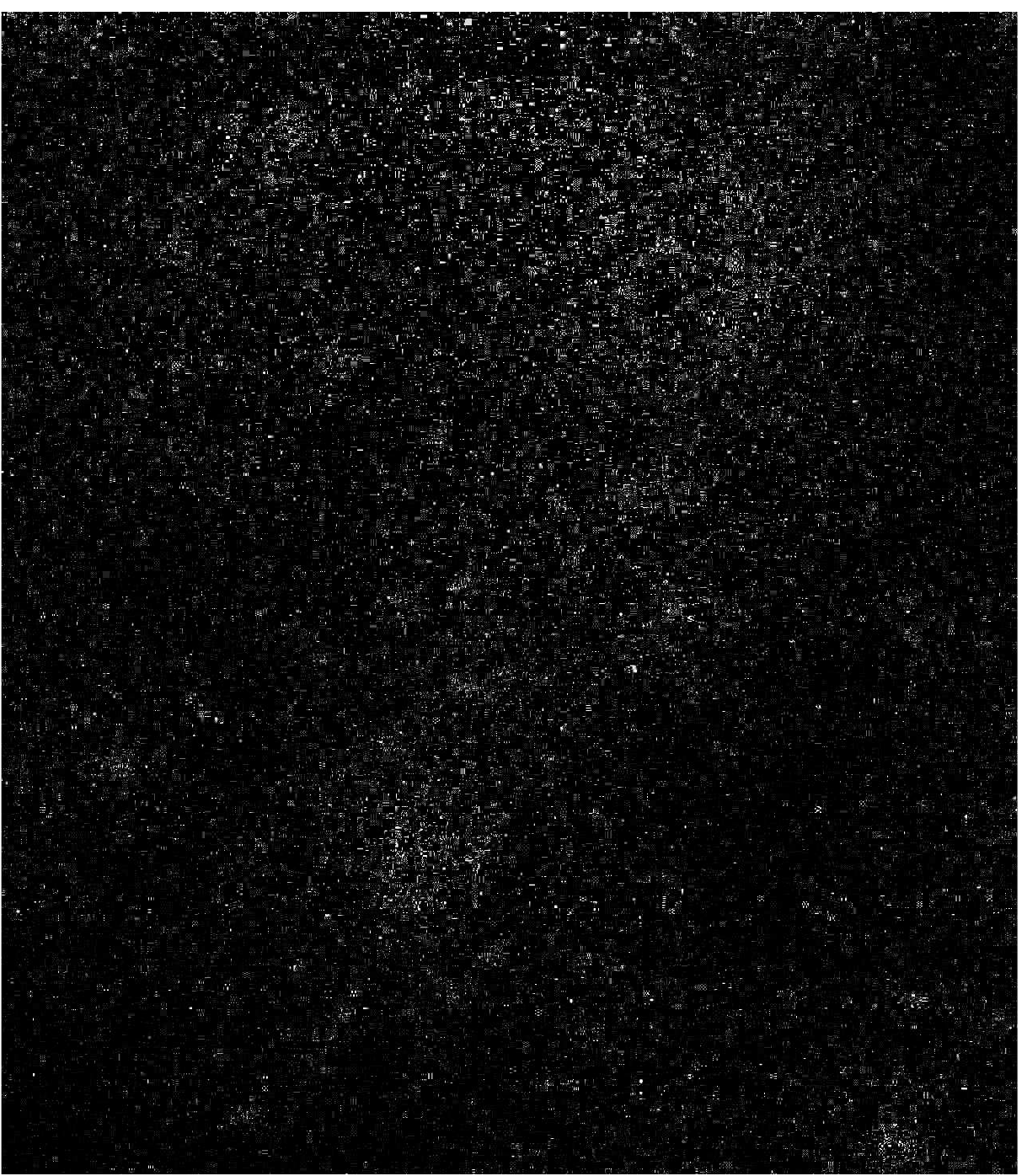}
    \includegraphics[width=0.32\textwidth]{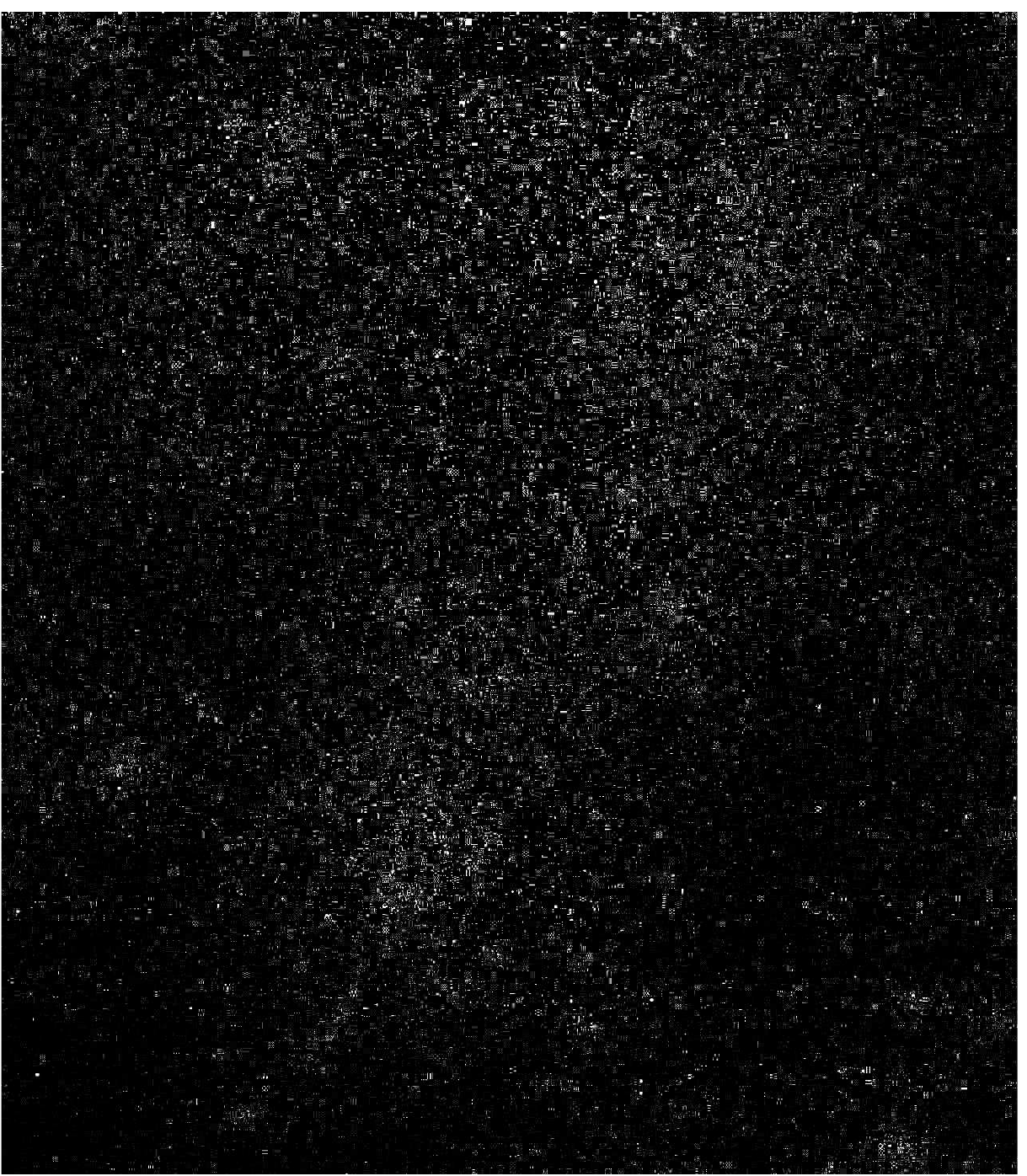}

    \includegraphics[width=0.32\textwidth]{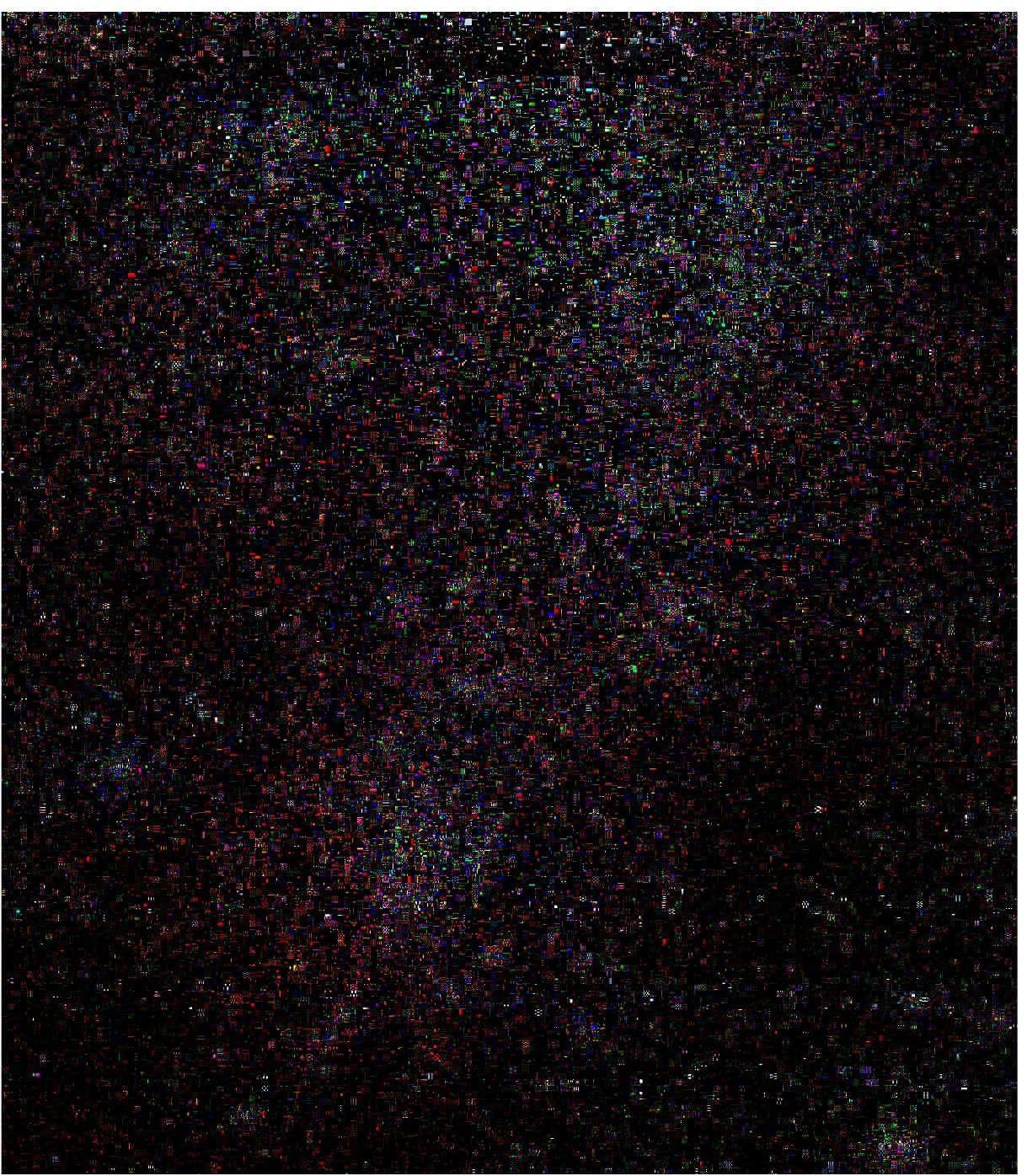}
  \end{center}
  \caption{\small{Unsuccessful attempt to expand the image 
  NGC 2264 of Fig.~2 using an incorrect key.}}
\end{figure}

Assuming now that the folded image is given to a partner stored in the original 16-bit RGB format, in order to recover the image the receiver should proceed as follows: first the header
information is read.  This is not encrypted and is required by the receiver to separate the 
image components $\Gt$ and $G$,
and to reconstruct the three independent folded images, 
displayed in the third row (from the top) of Fig. 2.

Now the process continues, as prescribed in Sec.~\ref{RP}, 
to recover the channels. 
The images in the fourth row of Fig.~2 depict the recovered channels using the  {\em{correct}} private key,
shown together as an RGB image in the last row. 
Because the authorized key is used, the recovery
was successful. 
Fig.~3 illustrates the identical process  using the 
{\em{incorrect}} private key. 
\begin{figure}
  \begin{center}
    \includegraphics[width=0.32\textwidth]{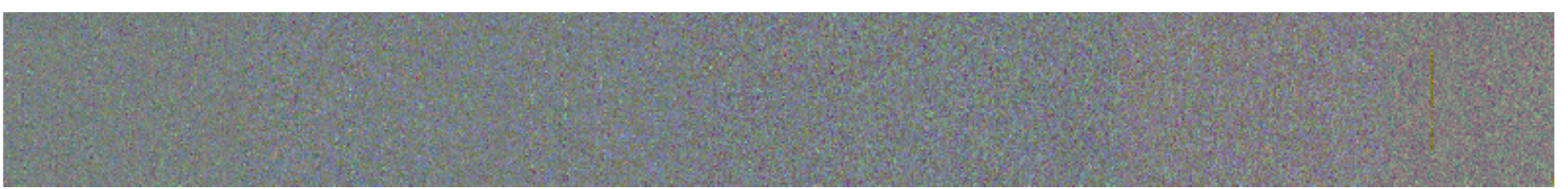}

    \includegraphics[width=0.32\textwidth]{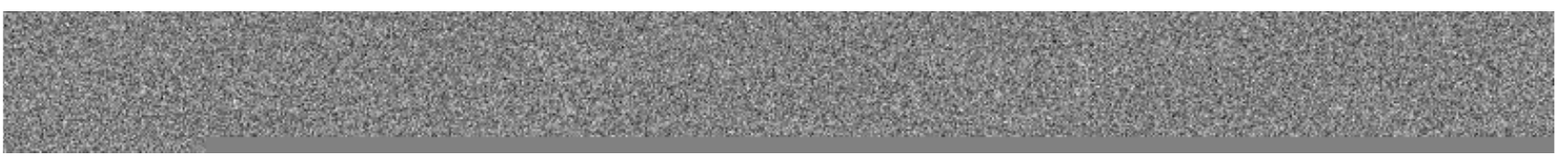}

    \includegraphics[width=0.32\textwidth]{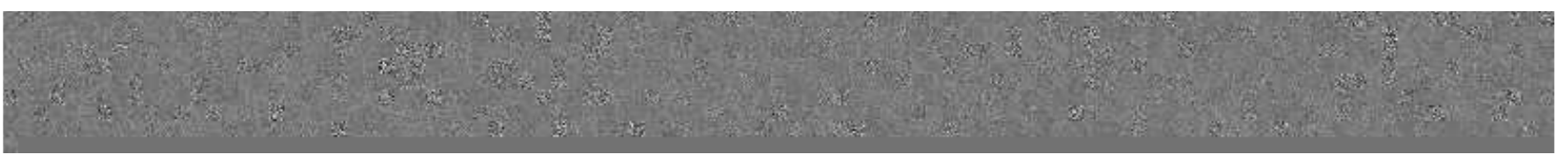}
    \includegraphics[width=0.32\textwidth]{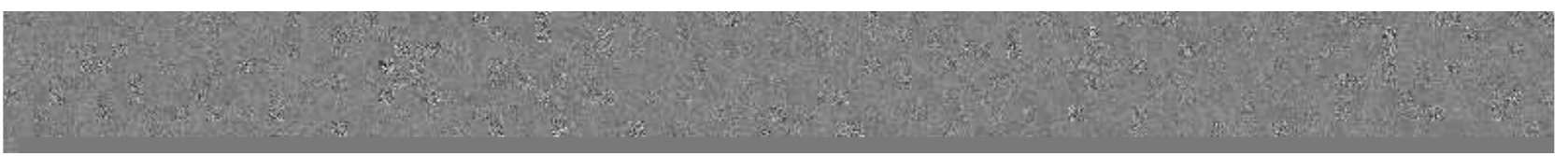}
    \includegraphics[width=0.32\textwidth]{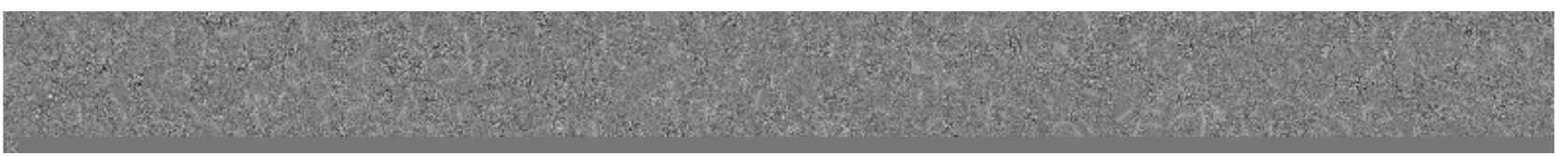}

    \includegraphics[width=0.32\textwidth]{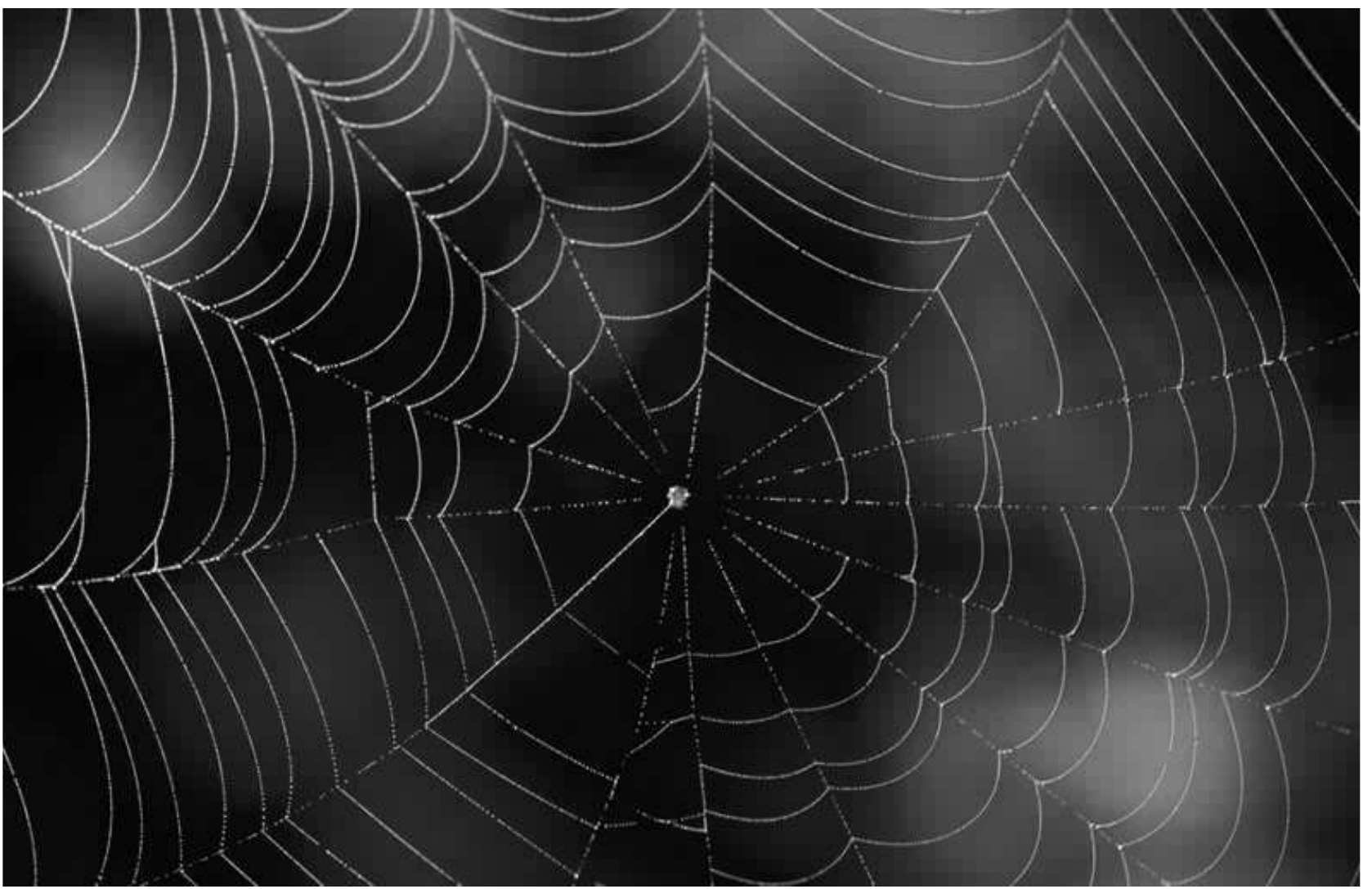}
    \includegraphics[width=0.32\textwidth]{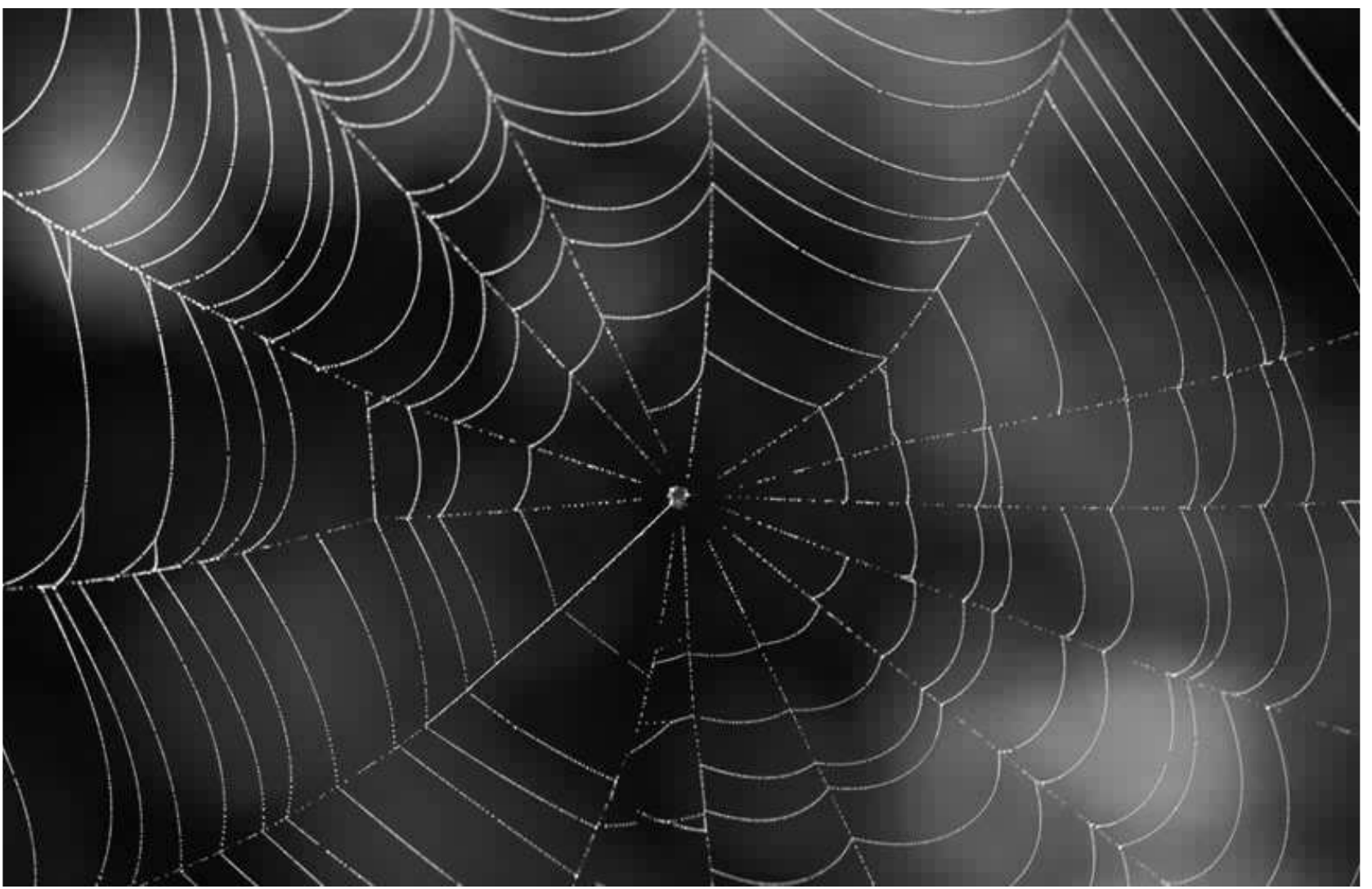}
    \includegraphics[width=0.32\textwidth]{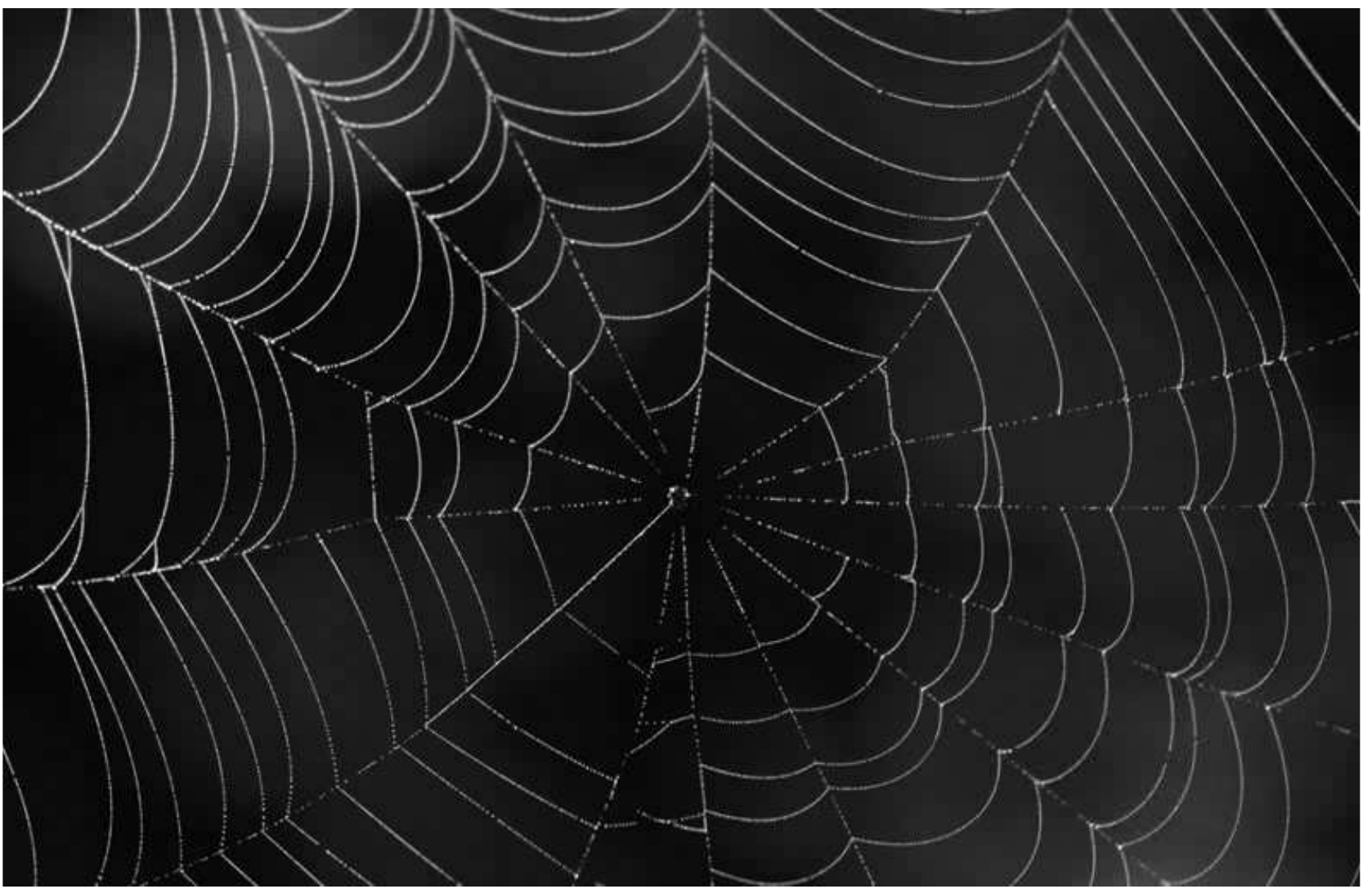}

    \includegraphics[width=0.32\textwidth]{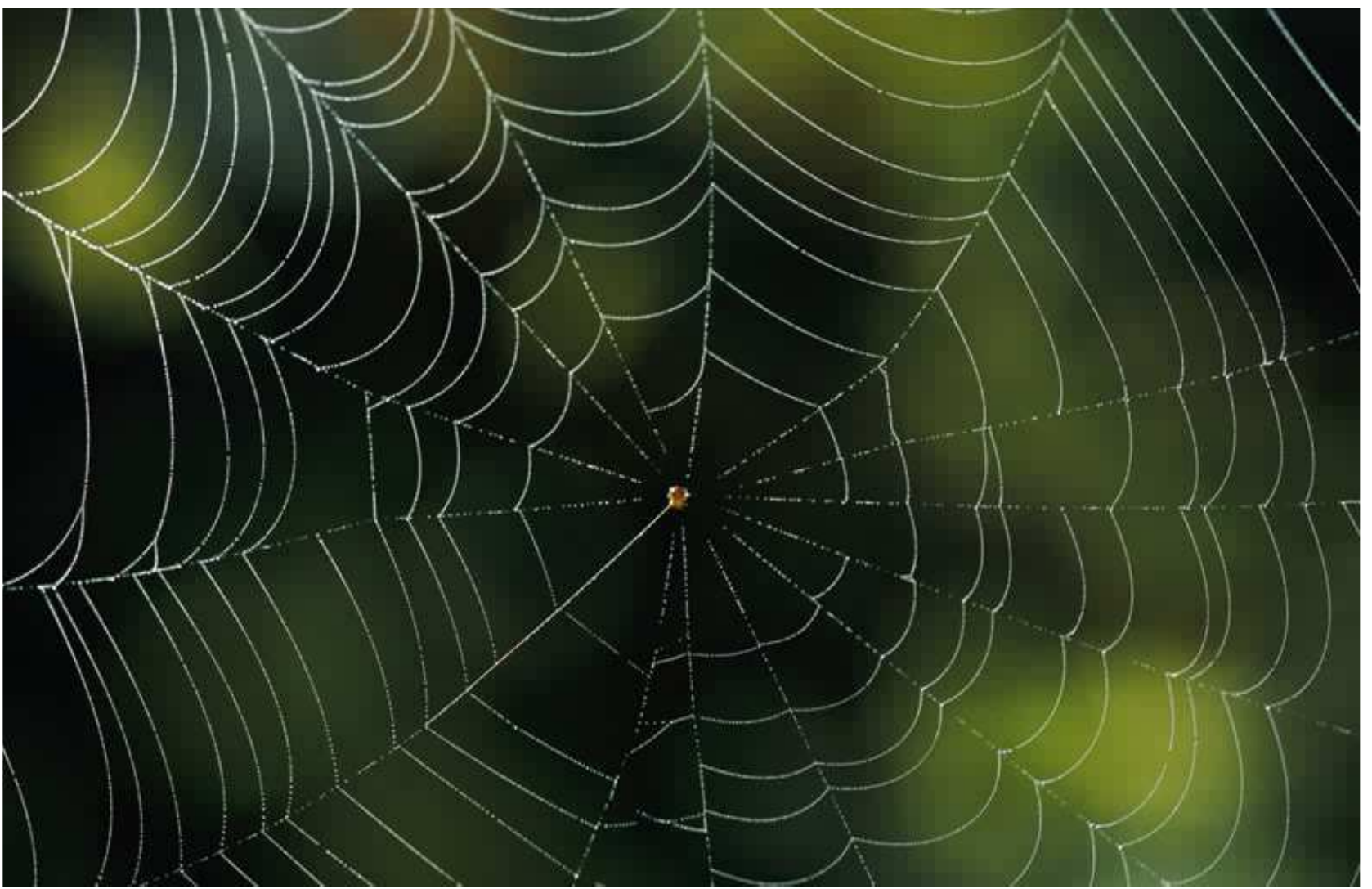}
  \end{center}
  \caption{\small{Same description as in Fig.~2 but the 
  image is a close up of a spider web in Australia. Courtesy of 
  National Geographic. Photograph by Darlyne Murawski \cite{NGWeb}.}}
\end{figure}
\begin{figure}
  \begin{center}
    \includegraphics[width=0.32\textwidth]{SCEIF_Images_Spider/spider_dict_folded}

    \includegraphics[width=0.32\textwidth]{SCEIF_Images_Spider/spider_dict_index_folded}

    \includegraphics[width=0.32\textwidth]{SCEIF_Images_Spider/spider_dict_folded_R}
    \includegraphics[width=0.32\textwidth]{SCEIF_Images_Spider/spider_dict_folded_G}
    \includegraphics[width=0.32\textwidth]{SCEIF_Images_Spider/spider_dict_folded_B}

    \includegraphics[width=0.32\textwidth]{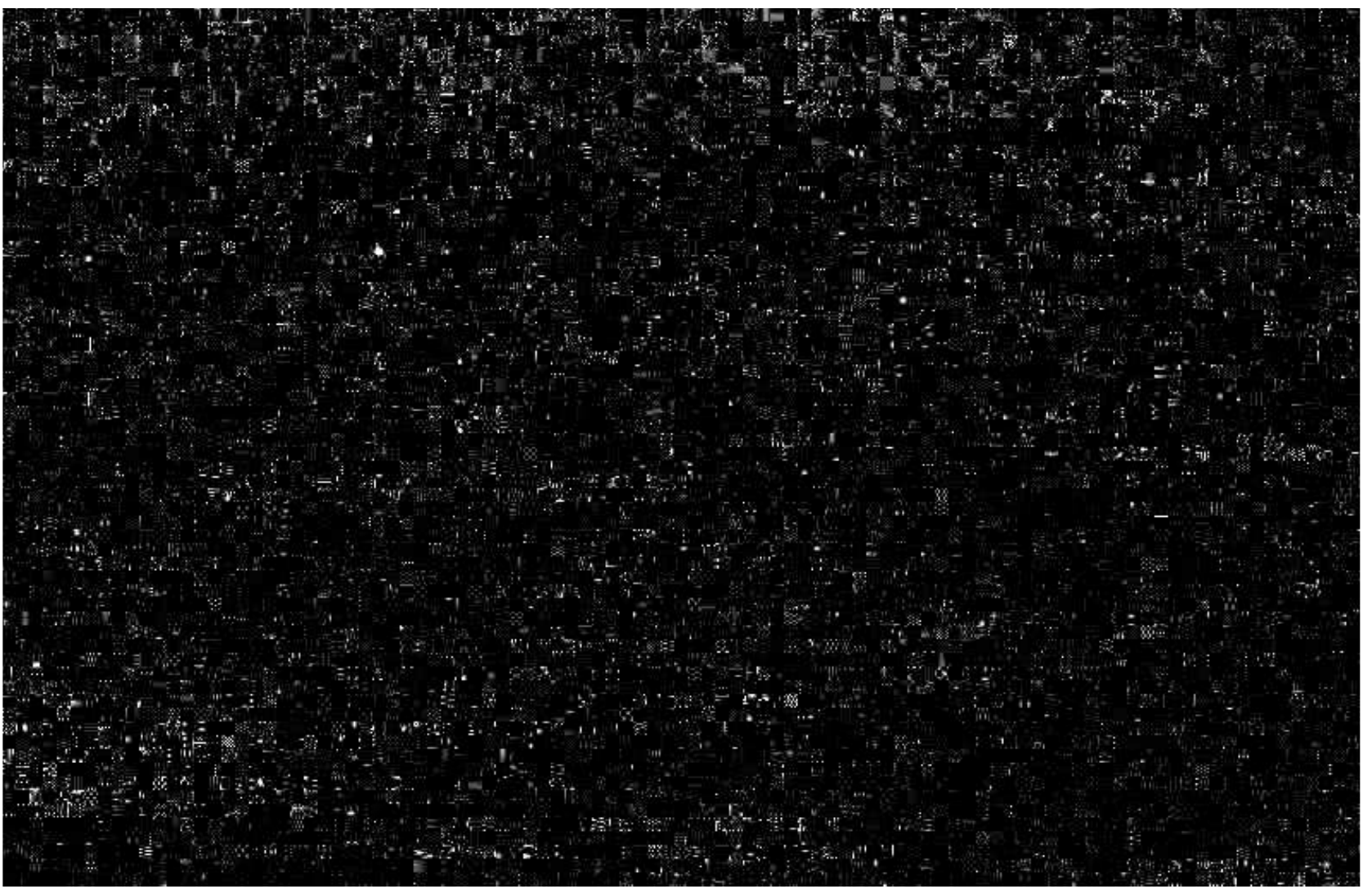}
    \includegraphics[width=0.32\textwidth]{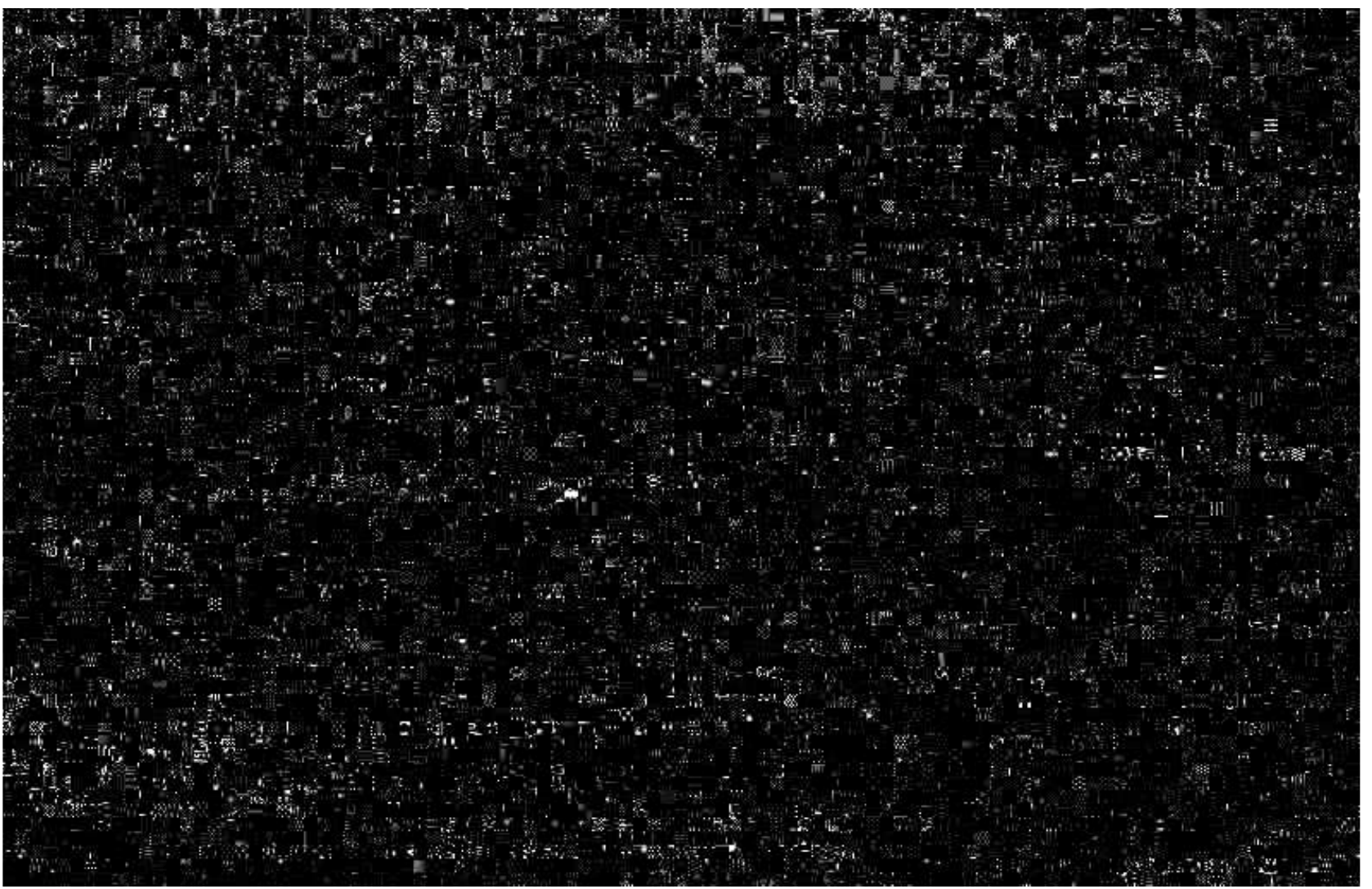}
    \includegraphics[width=0.32\textwidth]{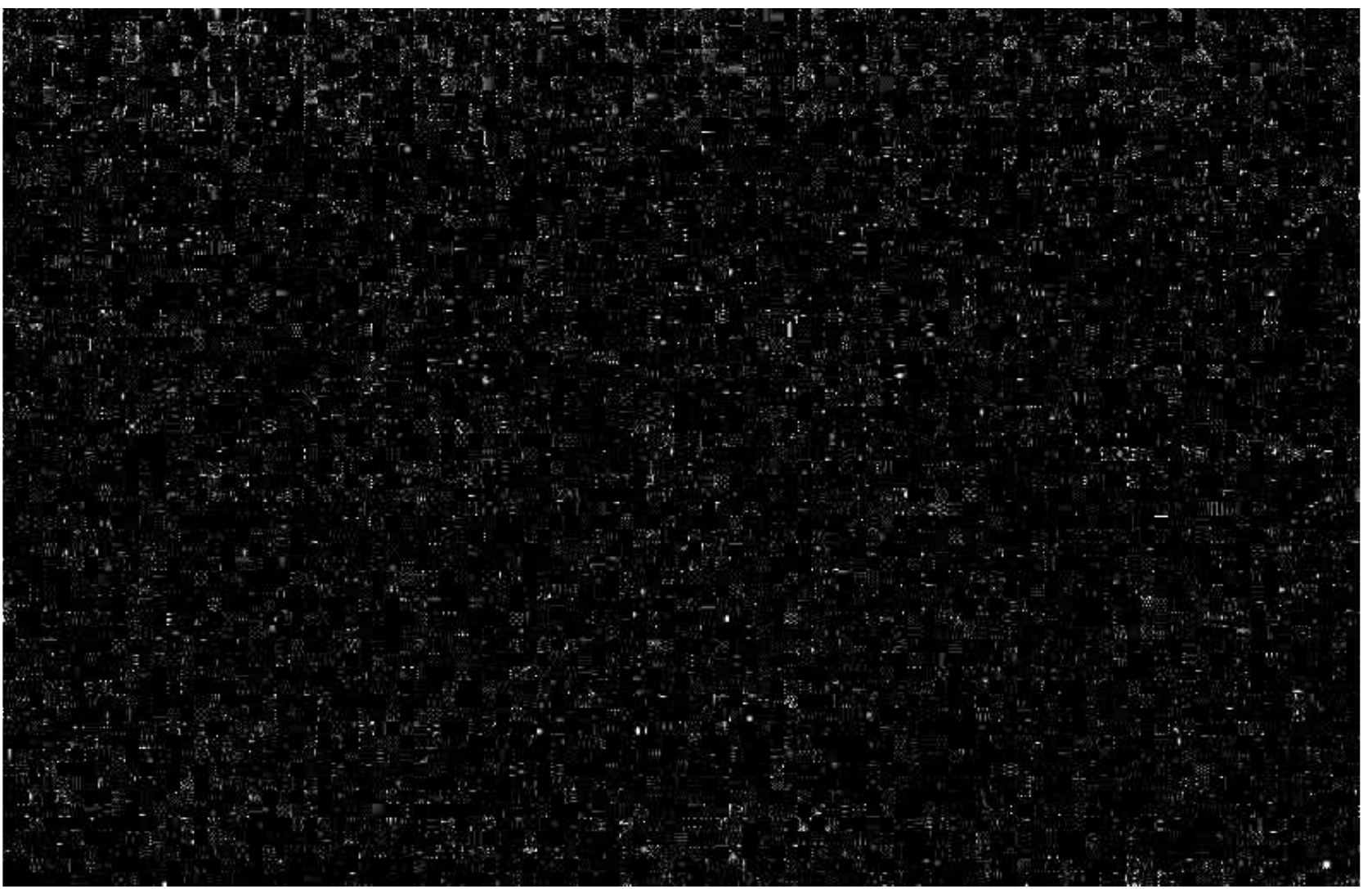}

    \includegraphics[width=0.32\textwidth]{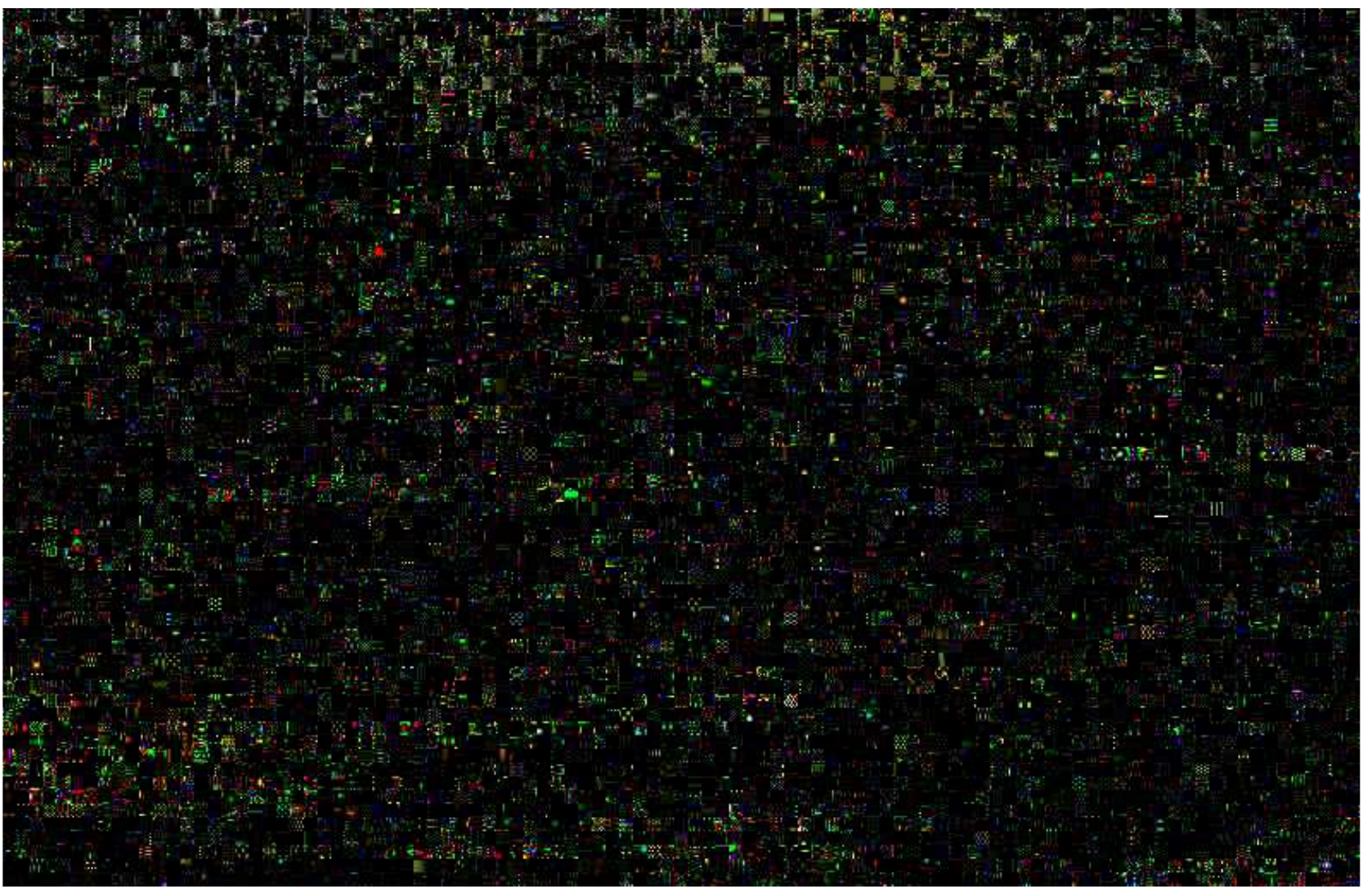}
  \end{center}
\caption{\small{Unsuccessful attempt to expand
  the spider web image of Fig.~4 using  an incorrect key.}}
\end{figure}
As a second example we proceed as before, but on 
a close up of the spider web photo, 
kindly rendered by National Geographic
\cite{NGWeb}. There is a difference from the 
previous case in that, instead of giving free access to 
the correct number of atoms per block $K_q,\,q=\Qt+1,\ldots,Q$, 
in this example those numbers 
are also hidden, together with the indices. The reason being that 
because of the contrast between the blocks containing the web 
and the rest of the blocks, those numbers give
some information about the image. Certainly, by knowing only
those numbers one can tell that the image has a very smooth 
background with some details only where the spider web is located. 
This gives some visual information that one may want to avoid by 
hiding those numbers.  

The folded image reduces the size 
of the original  spider web photo
($512 \times 792 \times 3$ pixels)
less than in the previous case ($89 \times 792 \times 3$ pixels)
because the SR is smaller: 7.95. 

For comparative purposes we have implemented the SCEIF 
method using DCT, which is also suitable for 
block processing. The implementation of the folding and encryption 
steps is exactly the same, the only difference is that the 
approximation can be performed by DCT, which is straightforward 
and faster than with the dictionary. 
However, since the sparsity achieved by 
DCT is lower (SR= 10.06 for the nebula image and SR = 4.23 for 
the spider web) the corresponding 
folded images are larger (see Fig.~6). In addition, because the 
processing time is dominated by the actual folding and 
expanding procedures, SCEIF implemented with the mixed dictionary 
is faster than with DCT (see Table \ref{table:ti}).
\begin{figure}
  \begin{center}
% original 1464 x 1280 x 3
    \includegraphics[width=0.32\textwidth]{SCEIF_Images_Astro/astro_dict_folded} % 120 x 1280 x 3
    \includegraphics[width=0.32\textwidth]{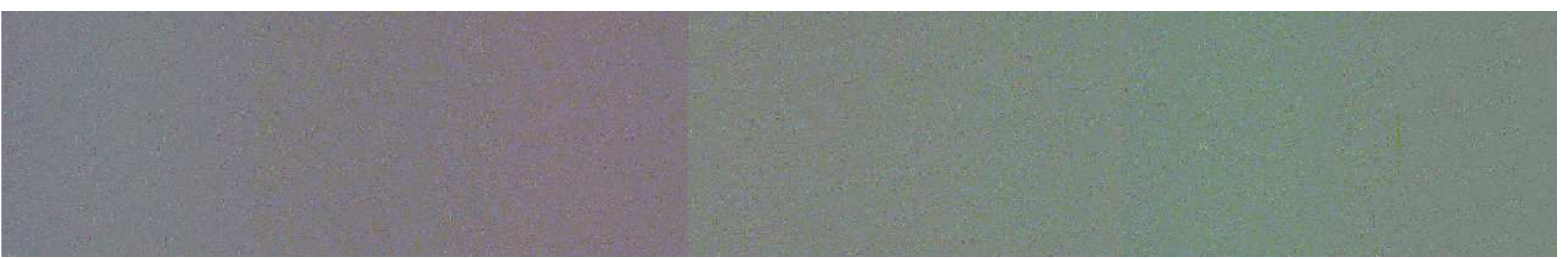} % 288 x 1280 x 3

% original 512 x 792 x 3
    \includegraphics[width=0.32\textwidth]{SCEIF_Images_Spider/spider_dict_folded} % 89 x 792 x 3
    \includegraphics[width=0.32\textwidth]{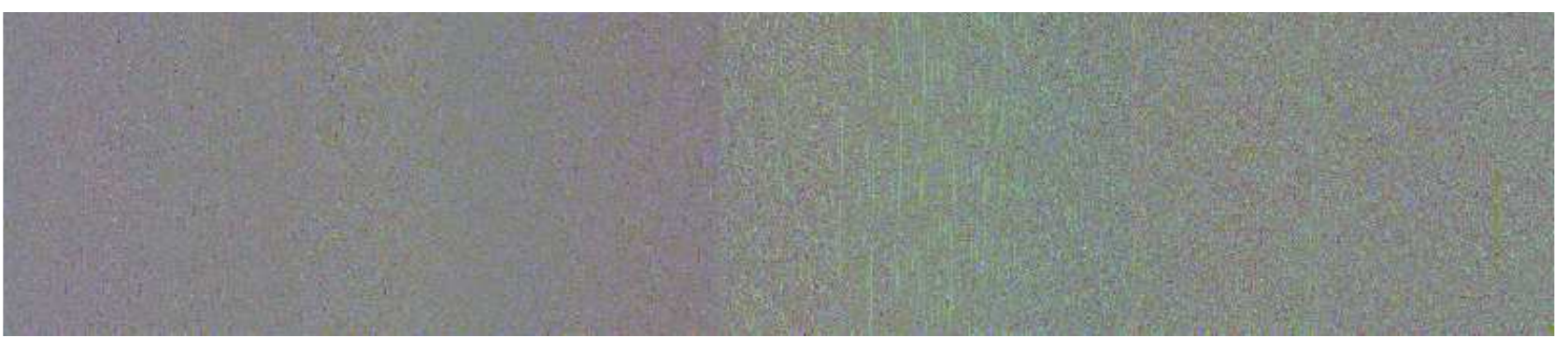} % 263 x 792 x 3
  \end{center}
  \caption{\small{The first picture in the top line
  is the folded image (size $120 \times 1280  \times 3$) of the 
  nebula NGC 2264 (size $1464 \times 1280  \times 3$) 
  with the proposed dictionary.
  The second picture in the top line  
  is the folded image (size $202 \times  1280  \times  3$) with 
  DCT. The pictures in the bottom line,  sizes 
  ($89 \times 792 \times  3$) and  ($165 \times 792 \times 3$), 
  are the folded images with the dictionary and 
  DCT, respectively, of the spider web image  
  (size $ 512 \times 792 \times 3$).}}
\end{figure}
% APPROXIMATION FOLDING AND EXPANDING TIMES
\begin{table}
\begin{center}
  \begin{tabular}{ll|l|l|l|l|}
    \cline{3-6}
    & & \multicolumn{4}{|c|}{Running times (in secs)} \\
    \cline{3-6}
    & & Approximation & Folding & Expanding & Total \\
    \cline{1-6}    
    \multicolumn{1}{|c|}{\multirow{2}{*}{Nebula}} & Dictionary & 10.9 & 10.7 & 13.7 & 35.3 \\
    \cline{2-6}
    \multicolumn{1}{|c|}{}
			   & DCT & \mbox{Disregarded} & 17.3 &  20.3 & 37.6 \\
    \cline{1-6}
    \multicolumn{1}{|c|}{\multirow{2}{*}{Spider web}} & Dictionary & 4.9 &
    4.7 & 5.6 & 15.2 \\
    \cline{2-6}
    \multicolumn{1}{|c|}{}
			   & DCT & \mbox{Disregarded} & 7.8 & 8.9 & 16.7\\
    \cline{1-6}
  \end{tabular}
    \caption{\small{Comparison of the folding and expanding times (average of five 
    independent runs)
    with the mixed dictionary and DCT. The 
    test was performed with MATLAB using 
    a 14'' laptop equipped with 2.8GHz processor and 3GB RAM. 
    As the implementation of the approximation with DCT 
    was not optimized, the approximation times are not included
    in the calculation of the total execution time with this approach.
    The approximation with the dictionary was realized using a MEX 
    file in C{\small{++}} for implementing OMP2DMl to approximate 
    the three channels simultaneously.}}
    \label{table:ti}
    \end{center}
\end{table}
\section{Quality and security issues}
\label{Ana}
Concerning the quality of the recovered image there are two
independent aspects to be discussed. One is the quality of 
the approximation, $I^K$, of the original image $I$ and 
the other is the quality of the recovery of $I^K$.

The security matters that will be discussed are restricted to
key sensitivity and resistance to plain text attack.
\subsection{Quality}
\label{qual}
The quality of the approximation, $I^K$, of the
image $I$ is to be decided beforehand. In the examples we have
considered high quality approximations. This is assessed
by two standard measures. One is the PSNR,
which is defined as
\be
\label{psnr}
\PSNR=10 \log_{10}\left(\frac{{(2^{l_{\bit}}-1)}^2}{\MSE}\right),
\ee
where $l_{\bit}$ is the number of bits used to represent the
intensity of the  pixels  and 
\be
\MSE=\frac{\sum_{z=1}^Z \| I_z - I^K_z\|_F^2}{Z\Nx \Ny},\nonumber
\ee
with $Z=1$ for a gray level image and $Z=3$ for an RGB image.

In the two numerical examples of Sec.~\ref{Expe}
the corresponding PSNR  is
high enough  (42.5 dB) to secure approximations of high
quality (with no visual degradation with respect to the
original image).
The other measure we have used to assess the quality  of the
approximate image is the
Mean Structure Similarity index (MSSIM) \cite{ssim}, which
for two identical images is equal to one.
The MSSIM index between the original image
and the approximation, in both examples of Sec.~\ref{Expe},
 is larger than 0.99. This value complements and confirms the
 quality indicated by the PSNR.

 Once the desired quality of the approximated image
 has been fixed, that approximation becomes
 the {\em plain text} image to be folded and encrypted.
 Thus, the next goal is to recover the
 approximate image  with high fidelity.
 The recovering would be `exact'  if not  
 for the quantization step which is introduced
 to store the folded image using integers.
 The present version of the proposed scheme works with images 
  stored using 16 bits per channel. At this
  precision, in both examples, the MSSIM index between
 the image recovered with the right key and the {\em plain
 text} image is equal to one. The PSNR between the authorized
 recovered image and
 the original image is
  identical to that between the plain text image and the
  original one.
  \subsection{Security}
  \label{sec}
The security of the encryption scheme we have adopted
relies on the random number generator.
The more reliable the random generator is
the safer the encryption procedure. Our implementation
  uses a simple 32-bit {\em{pseudo random number generator}} but,
  apart from the convenience of having it at hand, there is no
  reason for using that particular one.
%  On the contrary, the recommendation is, of course, to use a {\em{true random number generator}} if that is possible.

While the key space for the present implementation is $2^{32}$,
simply by making access to the order of orthogonalization private
(c.f. \eqref{Umat} and  \eqref{Umatt})
the key space would be expanded.

{\em{Key sensitivity}}:
 The high sensitivity against small variations in the private
 key is illustrated by Figures 3 and 5. The
 failed recovery  shown in those figures were attempted  using a
 key differing only by {\em{one}} digit with the correct one.
 The private key is 1234567891 and the
 tested key 1234567890.
 The PSNR between the plain text image and the recovered image
 with the wrong key is $10.8$dB for the image of Fig.~3  and 
 9.15 dB for the image of Fig.~5.
 This sensitivity was verified statistically by repeating 
 the experiment with 100 keys differing in only one digit from the correct
 key. The mean value of the resultant PSNR for the 
 nebula image is $10.68$dB with standard deviation 1.23. For 
 the spider web image the mean value PSNR is $8.6$dB with standard deviation
 1.41.

{\em{Prevention of plain text attacks:}}
In order to avoid repetitions of the encryption operators
(c.f. \eqref{Umat} and \eqref{Umatt})
the random arrays \eqref{Ymat} and \eqref{Ytmat}
should be guaranteed to be different every time the procedure is executed.
That is the role the public initialization $\seed$ plays  
at the folding step. The $\seed$ 
can be set automatically, for instance  as the date and time 
right before the vectors are generated. Thus, the 
nonlinearity of the operation  $\orth(\cdot)$  
prevents an attacker from inverting the system of equations \eqref{plain} 
and \eqref{plaint} using correctly 
decrypted plain text images.

Notice that the public $\seed$ ensures that even the identical
plain text image produces  a different cipher one.
In order to illustrate this feature we calculated the 
PSNR between two folded images encrypted with the same 
private key but different public seeds.
For the astronomical image the resulting PSNR
was 14.25 dB and for spider web 13.78 dB.\\

\section{Conclusions}
\label{Con}
The recently introduced EIF approach has been 
extended to SCEIF by introducing the following 
features:
\begin{itemize}
\item
The folding capacity of the approach has been improved by
considering  a new dictionary for the approximation.
\item
The approach is now self-contained. All that is required to 
recover the plain text image is the folded (cipher) image  and 
the private key. This is achieved by enlarging the 
folded image creating ad hoc blocks to place the indexes 
of those dictionary's elements  participating 
in the image approximation (plain text image).
\item
The implementation has also been extended from gray level to 
color images. 
\end{itemize}
The success of the approach is based on 
two fundamental and related features: One is the possibility of 
reducing the data dimensionality by a powerful highly non linear
transformation. 
The other is the possibility of implementing  the approach in 
an affordable period of time.
The proposed dictionary plays a central role in ensuring 
both features, by allowing for processing obeying a {\em{scaling law}}. 
Certainly, the fact that the approximation of a large image 
can be realized  by dividing it into small blocks 
is the key of the current effective implementation. 
It should be emphasized that the numerical examples have been 
realized on a small laptop in MATLAB environment. 
Simply by implementing the method in a programming language, 
such as C or Fortran, the
folding and expanding times given in Table \ref{table:ti}
could be reduced by up to tenfold. In addition, 
there is room for straightforward implementation by parallel 
computing if those resources are available.
\subsection*{Final Remarks}
\begin{itemize}
\item
The scope of SCEIF is to fold an image in
encrypted form. The size of the astronomical image is
reduced 12.2-fold (pixel wise) and the spider web 5.75-fold.
We are not considering here any further
compression stage, which could imply to convert the folded image into
a bit stream. It should be stressed that, in order to do that,
the encoding technique should be especially conceived to deal with
the type of data that SCEIF generates by folding the image.
\item
The simple symmetric key encryption procedure considered
here leaves room for straightforward improvement, e.g.,

a)The key space could be extended by the orthogonalization operation.
In the  present version
the orthogonalization step (c.f. \eqref{Umat} and \eqref{Umatt})
is assumed to be completely known.  
However, simply by making access to the order of orthogonalization 
private, the key space would be expanded.

b)The other possibility that can be foreseen, to
strengthen the security of the
proposed encryption scheme, is to further scramble
the folded image using a
chaos-based encryption algorithm. Considering the
security flaws affecting  some of those algorithms
\cite{ARA08,ALA10,PZ10,AAA11,SCY10}, it becomes noticeable that our
approach could benefit those techniques in a 
twofold manner:
i)providing a way of reducing the image size, and
ii)enhancing the security of the algorithms.
\end{itemize}
For the above reasons the proposed SCEIF approach 
appears in our mind a very exciting possibility. 
We feel confident that it will stimulate further 
work in this direction. 
\subsection*{Acknowledgements}
Support from EPSRC, UK, grant  (EP$/$D062632$/$1) is acknowledged.
We are grateful to the  European Southern Observatory 
and National Geographic for authorization to use 
their images for illustration of the approach. Dedicated 
open source software  for 
implementation of the approach is available at \cite{Webpage2},  
section SCEIFS.
%\newpage
\appendix
\subsection*{A. Construction of Matrices $B^k_n,\,n=1,\ldots,k$ 
(c.f.\eqref{Bcoe})}
\renewcommand{\theequation}{A.\arabic{equation}}
\setcounter{equation}{0}
For $z=1,2,3$ the coefficients $c_{n}^{z},\,n=1,\ldots,k$ 
in \eqref{ompml}  should be determined in  such  a way that 
$\|R_{z}^{k}\|_\F$ is minimum for each $z$.
This is ensured by requesting that $R_{z}^{k}= I_{z}- \op_{\V_k} I_{z},\,
z=1,2,3$,
where $\op_{\V_k}$ is the orthogonal
projection operator
onto $\V_k=\Spann\{\dx_{\ell^x_n} \otimes \dy_{\ell^y_n} \}_{n=1}^k$.
The required  representation of $\hat{P}_{\V_k}$ is of the form
$\hat{P}_{\V_k} I  = \sum_{n=1}^k A_n \la B_n^k, I \ra_F $,
where each $A_n \in \R^{\Nxy}$ is  an array with the
selected atoms $A_n= \dx_{\ell^x_n} \otimes \dy_{\ell^y_n}$
and  $B_n^k,\,n=1,\ldots,k$  the
 concomitant reciprocal matrices. These
 are the unique elements
 of $\R^{\Nxy}$  satisfying the conditions:
 \begin{itemize}
 \item
 [i)]$\la A_n, B_m^k \ra_\F=\delta_{n,m}= \begin{cases}
 1 & \mbox{if}\, n=m\\
 0 & \mbox{if}\, n\neq m.
 \end{cases}$
\item
[ii)]${\V_k}= \Spann\{B_n^k\}_{n=1}^k.$
\end{itemize}
Such matrices can be adaptively constructed through the recursion formula
\cite{RNL02}:
\be
 \begin{split}
 B_n^{k+1}&= B_n^k - B_{k+1}^{k+1}\la A_{k+1}, B_n^k \ra_\F,\quad n=1,\ldots,k,\\
 \text{where}\\
 B_{k+1}^{k+1}&= W_{k+1}/\| W_{k+1}\|_\F^2,\,\,
 \text{with}\,\,
 W_1=A_1 \,\, \text{and} \,\,
 W_{k+1}= A_{k+1} - \sum_{n=1}^k \frac{W_n}
 {\|W_n\|_\F^2} \la W_n, A_{k+1}\ra_\F.
 \end{split}
 \ee
For numerical accuracy in
$W_{n},\,n=1,\ldots,k+1$ at
least one re-orthogonalization step
is usually needed. It implies that one needs to 
recalculate these matrices as
\be
\label{GS}
W_{k+1}= W_{k+1}- \sum_{n=1}^k \frac{W_n}{\|W_n\|_\F^2}
\la W_n , W_{k+1}\ra_\F.
\ee
With matrices $B_n^k,\,n=1,\ldots,k$ constructed as above
the required coefficients
in \eqref{atomml} are obtained, for $z=1,2,3$, from the
inner products
$$c^{z}_n= \la B_n^k, I_{z} \ra_\F,\, n=1,\ldots,k.$$ 
\bibliographystyle{elsart-num}
\bibliography{revbib}
\end{document}